\documentclass{aa}

\usepackage{graphicx}
\usepackage{txfonts}
\usepackage{natbib}
\usepackage[breaklinks=true, colorlinks=true, allcolors=blue]{hyperref}
\usepackage{amsmath}	
\usepackage{amssymb}	
\usepackage{multirow}
\usepackage{longtable}

\usepackage{multicol}
\usepackage{bm}

\begin{document}

   \title{An $nl$-model with full radiative transfer treatment for level populations of hydrogen atoms in spherically symmetric H II region}

   \author{F.-Y. Zhu\inst{1},
          J. Wang\inst{2},
          Y. Qiu\inst{1},
          Q.-F. Zhu\inst{3},
          \and
          D. Quan\inst{1}
          }

   \institute{Research Center for Astronomical Computing, Zhejiang Laboratory, Hangzhou 311100, PR China\\
              \email{zhufy@zhejianglab.com; donghui.quan@zhejianglab.com}
         \and
             School of Physical Science and Technology, Guangxi University, Nanning 530004, PR China \\
             \email{junzhiwang@gxu.edu.cn}
         \and
             Department of Astronomy, University of Science and Technology of China, Hefei, 230026, PR China
             }

 \date{Received xx; accepted xxx}

 \titlerunning{An $nl$-model of RRLs with full radiative transfer \textbf{treatment}}
 \authorrunning{Zhu et al.}


  \abstract
   {The radiation field due to hydrogen recombination lines and continuum emission could significantly affect the hydrogen level populations in ultra- and hyper-compact H II regions. The escape probability approximation was used to estimate the effect of radiation field in previous models for calculating hydrogen level populations. But the reliability of this approximation has not been systematically studied.}
   {In this work, the appropriate ranges of the previous models with the escape probability approximation and without the effects of radiation field are investigated. A new model is created to simulate both the integrated characteristics and the spatially resolved diagnostics of hydrogen recombination lines throughout H II regions.}
   {A new $nl$ model with full radiative transfer treatment of radiation field due to hydrogen recombination lines and continuum emission is developed to calculate the hydrogen level populations and hydrogen recombination lines. The level populations and the corresponding hydrogen recombination line intensities simulated by the new model and previous models are compared.}
   {The applicability and the valid parameter ranges of the previous models are studied. Radiation fields exhibit negligible effects on level populations in both classical and UC H II regions. With the modified escape probability, the model with the escape probability approximation is suitable for most cases of HC H II regions. The improved new model performs better in the HC H II region with extremely high emission measure (EM). To address the high computational costs inherent in numerical models, a precise machine-learning model is trained to enable rapid estimation of hydrogen level populations and associated hydrogen recombination lines.}
   {}


   \maketitle
   
%

\section{Introduction} \label{sec:intro}

Massive stars are important for the evolution of galaxies because of their powerful feedback including outflows, stellar wind, and ultraviolet radiation. They are always born in dense molecular clouds. When a massive star evolves into the main-sequence stage, the surrounding molecular gas can be ionized due to the extreme-ultraviolet (EUV, $h\nu\geq$ 13.6 eV) radiation from the ionizing massive star. Then an ultra-/hyper-compact (U/HC) H II region is produced. Ionized gas consisting of ionized hydrogen and free electrons is the main component in the H II region. And the studies of the ionized gas in U/HC H II regions are useful to understand the feedback and properties of the central massive stars. Observations of ionized gas are also used to evaluate star formation activities in some molecular clouds in the Galaxy and even in external galaxies \citep{zha22,zha23}.

Hydrogen radio recombination lines (RRLs) and continuum emission are the main tools to investigate the properties of ionized gas. The line-to-continuum ratio is commonly used to estimate the electron temperature of the ionized gas under the assumption of local thermodynamic equilibrium (LTE) conditions \citep{chu75,sha83}. However, the electron temperatures of some H II regions in the Galaxy estimated from a line-to-continuum ratio in centimeter wavelengths under the LTE assumption were found significantly lower than those estimated from forbidden optical lines \citep{sor66,mez67,die67}. This is produced by significant stimulated emissions of RRLs because the level populations of hydrogen atoms in H II regions depart from those under the LTE assumption \citep{gor02}. Then, the accurate calculations of departure coefficients are important to estimate the electron temperatures of ionized gas from the intensities of hydrogen RRLs and radio continuum.

The departure coefficients of the hydrogen level populations from those under the LTE assumption have already been investigated for decades \citep{bak38,sej69}. In the early years, only the transitions between different principal quantum states were considered \citep{bro70,bur76,wal90}. This kind of method is called $n$-model. Several $nl$-models including angular momentum changing transitions were developed later \citep{bro71,hum87,sto95}. Although both of $n$- and $nl$-models were applied for the hydrogen RRLs with obvious maser or significant stimulated emission \citep{aff94,bae13}, these models are all based on the assumption that radiation of RRLs and continuum emission produced by H II regions have no significant influence on the level populations. However, this assumption may not be appropriate for some ultra/hyper-compact HII regions, especially for those with strong masers of hydrogen RRLs and optically thick continuum emission \citep{pro20}. In order to solve this problem, the effects of continuum emission and RRLs with a box profile were incorporated into an $nl$-model \citep{pro18,pro20}. This kind of $nl$-model was recently improved by \citet{zhu22} with realistic profiles instead of box profiles.

In order to calculate the effects of line and continuum radiation fields ($J_\nu$) on the level populations of hydrogen atoms in H II regions, the radiation should be calculated non-locally in principle \citep{zhu22}. For time saving, simplifying assumptions were used in the previous works. The escape probability approximation (EPA) was adopted which implies the source function ($S_\nu$) and mean intensity of the incident radiation field are homogeneous throughout every part of an H II region \citep{pro20,zhu22}. However, the escape probability approximation is not from first principles. This could lead to some deviations. It is necessary to check the reliability of this approximation. In some previous works, the results calculated under the approximation were compared with those derived from full radiative transfer treatment \citep{dum03,nes16}. But the targets in these works were not hydrogen RRLs. So in order to assess the reliability of the escape probability approximation for RRLs, it is important to develop an improved model for RRLs with full radiative transfer treatment. In addition, this kind of model is also essential if we want to know the detailed properties of hydrogen RRLs and the hydrogen level populations at different locations in an H II region.

In this work, we develop an $nl$-model with full radiative transfer treatment to derive the level populations of hydrogen atoms and properties of hydrogen RRLs for spherical U/HC H II regions. Compared with the new model, the accuracy of the previous model with the escape probability approximation is assessed. The application conditions of the approximation are investigated. The influence of spatial variation in hydrogen level populations within U/HC H II regions on profiles and intensities of hydrogen RRLs is also studied. In addition, based on an extensive dataset generated by this new model, a machine-learning model is made using a random forest regressor in order to avoid the time-consuming calculations using the numerical models. 

The organization of the current work is as follows. The method of the improved $nl$-model with full radiative transfer treatment of the hydrogen RRLs and continuum emission is explained in Section \ref{sec:method}. In Section \ref{sec:result}, the results and the discussions of the new model that is used to study the hydrogen level populations and properties of hydrogen RRLs for U/HC H II regions are presented. A machine-learning model based on the results of the new model is also introduced. In Section \ref{sec:conclusion}, the summary and the conclusions are given.

\section{Numerical Method} \label{sec:method}

\subsection{The distributions of electron temperature and density in spherical H II regions}

The improved current $nl$-model in this work is applied to spherically symmetric H II regions. The electron temperature ($T_e$) is uniform in the whole H II region. The electron density ($n_e$) can vary with the distance to the central ionizing star. The radial velocity of ionized gas can also vary with the radius while the velocity in all other directions is always zero. In addition, the micro-turbulent velocity is assumed to be constant within an H II region.

\subsection{The calculation of departure coefficients in Case B}

The calculations of hydrogen level populations and departure coefficients are most crucial to simulate the intensities of hydrogen recombination lines and continuum emission. The level population equations can be thought to be composed of a series of linear equations \citep{sal17,pro18}. These equations can be written in a matrix form as
\begin{equation}
\textbf{A}\cdot\textbf{b}=\textbf{y}~~~,
\end{equation}
where the elements of matrix \textbf{A} and vector \textbf{y} refer to the processes of transitions among the free and bound states with different principal quantum number $n$ and angular momentum number $l$. The vector \textbf{b} consists of partial departure coefficients ($b_{nl}$). If the effects of absorption and stimulated emission due to radiation fields are not significant, the values of the $b_{nl}$ can be directly calculated by solving the matrix equation \citep{zhu22}. In the current paper, this kind of models such as \citet{sto95} and \citet{zhu19} are called Case B models.

\subsection{The model under the escape probability approximation}

When the effects of radiation fields on departure coefficients are negligible, the $b_{nl}$ can be directly obtained by solving the matrix equation mentioned above. If the radiation fields of hydrogen recombination lines and continuum emission are important, an iteration method is necessary because of the interaction between the incident radiation and the departure coefficients \citep{zhu22}. In principle, the intensities of incident radiation should be computed non-locally. However, this may cause the calculations to be time-consuming. So in order to make the calculation tractable, the escape probability approximation was used in previous works \citep{pro20,zhu22}. In these models under the approximation, the mean intensity of the incident radiation field is 
\begin{equation}
J_\nu=S_\nu(1-e^{-\tau_\nu})~~~,
\end{equation}
where $\tau_\nu$ is the optical depth at the frequency $\nu$ and $e^{-\tau_\nu}$ is the escape probability. Following \citet{pro20}, in our previous work \citep{zhu22}, the value of $\tau_\nu$ is the optical depth given by
\begin{equation}
\tau_\nu=\kappa_\nu D~~~,
\end{equation}
where $\kappa_\nu$ is the absorption coefficient at the frequency $\nu$. $D$ is the thickness of a homogeneous medium or the diameter of a uniform spherical H II region.  In this paper, this kind of models are called the EPA models.

\subsection{Radiative transfer in the improved model with full radiative transfer treatment}

\begin{figure}
\begin{center}
\includegraphics[scale=0.6]{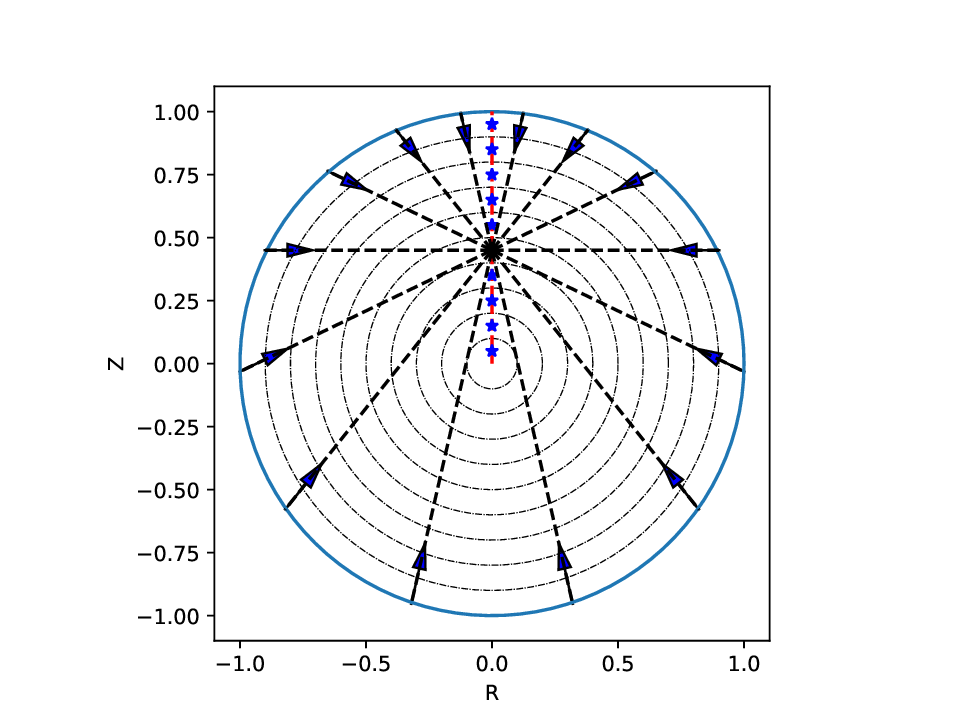}
\caption{The schematic diagram of radiative transfer of hydrogen RRLs and continuum emission in the new model. The large blue circle is the boundary of a spherical H II region. The blue asterisks are the grid points where the mean incident radiation field needs to be calculated. The black dashed lines are the radiation transfer paths in different directions. The arrows indicate the direction of propagation of RRLs and continuum emission. R and Z are the normalized radial and axial coordinates in the axisymmetric coordinate system.}\label{fig:transfer}
\end{center}
\end{figure}

As mentioned in Section \ref{sec:intro}, the EPA model is not from first principles. It is necessary to assess the reliability of this kind of model. In addition, EPA methods can only give global properties of hydrogen RRLs from an H II region. The exact details of hydrogen RRLs within the H II region cannot be provided \citep{pro20}. So an improved model with full radiative transfer treatment about hydrogen RRLs and continuum emission is essential. In this work, the properties of continuum emission and hydrogen RRLs are calculated non-locally. For calculating the intensities and velocity/frequency distribution of incident radiation fields in a certain position, the radiative transfers of RRLs and continuum emission are calculated in a series of directions as shown in Figure \ref{fig:transfer}. To estimate the mean incident radiation field in a position of a spherically symmetric H II region, the radiative transfer should be calculated in a two-dimensional axisymmetric coordinate system.

For the ionized gas in a certain position, the continuum absorption coefficient $\kappa_{\nu,\textrm{C}}$ given by \citet{ost61} and \citet{gor02} with the Rayleigh-Jeans approximation ($h\nu \ll kT_e$) is calculated numerically as
\begin{equation}
\kappa_{\nu,\textrm{C}}=9.770\times10^{-3}\frac{n_en_i}{\nu^2T_e^{1.5}}[17.72+ln\frac{T_e^{1.5}}{\nu}]~~~,
\end{equation}
where $\nu$ is the frequency in Hz. $n_i$ and $n_e$ are the number densities of ions and electrons in units of cm$^{-3}$, respectively. $T_e$ is the electron temperature in units of $K$, and $\kappa_{\nu,C}$ is in units of cm$^{-1}$. As written in \citet{pet12}, the continuum emissivity $j_{\nu,\textrm{C}}$ is
\begin{equation}
j_{\nu,\textrm{c}}=B_\nu(T_e)\kappa_{\nu,\textrm{C}}~~~,
\end{equation}
where $B_\nu(T_e)$ is the intensity of a blackbody of temperature $T_e$ at frequency $\nu$. The line absorption coefficient for the transition from lower level $n$ to upper level $m$ is
\begin{equation}
\kappa_{\nu,\textrm{L}}=\frac{h\nu}{4\pi}\sum_{l=0}^{n-1}\sum_{l'=l\pm1}(N_{nl}B_{nlml'}-N_{ml'}B_{ml'nl})\phi_\nu~~~,
\label{equ:cross}
\end{equation}
where $N_{nl}$ is the number density with the principal quantum number $n$ and angular momentum number $l$. $B_{nlml'}$ are the Einstein coefficients for absorption and stimulated emissions of the transitions. They can be calculated from the corresponding spontaneous Einstein coefficients $A_{ml'nl}$ \citep{bro71,pro20,zhu22}. The line profile function $\phi_\nu$ is calculated by considering the thermal,  turbulent, and pressure broadenings \citep{pet12}. The populations of hydrogen atoms in state $n$ and $nl$ are respectively \citep{gor02}
\begin{equation}
N_n=b_n\frac{n_en_i}{T_e^{1.5}}\frac{n^2h^3}{(2\pi m_ek)^{1.5}}\exp(\frac{E_n}{kT_e})~~~,
\end{equation}
and
\begin{equation}
N_{nl}=b_{nl}\frac{n_en_i}{T_e^{1.5}}\frac{(2l+1)h^3}{(2\pi m_ek)^{1.5}}\exp(\frac{E_n}{kT_e}) ~~~,
\end{equation}
where $k$ is the Boltzmann constant, $h$ is the Planck constant, $E_n$ is the energy of level $n$ below the continuum. $m_e$ is the electron mass. The line emission coefficient is written as
\begin{equation}
j_{\nu,\textrm{L}}=\frac{h\nu}{4\pi}\phi_\nu\sum_{l'=0}^{m-1}\sum_{l=l'\pm1}N_{ml'}A_{ml'nl}~~~.
\label{equ:emission}
\end{equation}

The radiative transfer equation is given by \citet{pet12} as
\begin{equation}
\frac{dI_\nu}{dx}=(j_{\nu,\textrm{C}}+j_{\nu,\textrm{L}})-(\kappa_{\nu,\textrm{C}}+\kappa_{\nu,\textrm{L}})I_\nu~~~,
\end{equation}
where $x$ is the spatial position. The optical depth is
\begin{equation}
\tau_\nu(x)=\tau_{\nu,\textrm{C}}(x)+\tau_{\nu,\textrm{L}}=\int_0^x\kappa_{\nu,\textrm{C}}(x')+\kappa_{\nu,\textrm{L}}(x')dx'~~~.
\end{equation}
And the source function $S_\nu$ is
\begin{equation}
S_\nu=\frac{j_{\nu,\textrm{L}}+j_{\nu,\textrm{C}}}{\kappa_{\nu,\textrm{L}}+\kappa_{\nu,\textrm{C}}}~~~.
\end{equation}

For continuum emission, the radiative transfer equation is
\begin{equation}
\frac{dI_{\nu,\textrm{C}}}{dx}=j_{\nu,\textrm{C}}-\kappa_{\nu,\textrm{C}}I_{\nu,\textrm{C}}~~~.
\end{equation}
The corresponding source function $S_{\nu,\textrm{C}}$ is equal to $B_\nu(T_e)$. The total intensity of line and continuum emission is written as \citep{pet12}
\begin{equation}
I_\nu(\tau_\nu)=I_\nu(0)e^{-\tau_\nu}+e^{-\tau_\nu}\int_0^{\tau_\nu} e^{\tau'_\nu}S_\nu(\tau'_\nu)d\tau'_\nu~~~,
\end{equation}
where $I_\nu(0)$ is the background intensity. Then the mean intensity of the incident RRLs and continuum emissions for the ionized gas in a certain position is given by
\begin{equation}
J_\nu=\frac{1}{4\pi}\int_0^{4\pi}I_{\nu}d\Omega=\frac{1}{2}\int_0^{\pi}I_{\nu}sin\theta d\theta~~~,
\end{equation}
where $\Omega$ is the solid angle, and $\theta$ is the angle between the direction of the radiative transfer path and the \textbf{z-axis} in the axisymmetric coordinate system. The mean radiation field is 
\begin{equation}
\bar{J}=\int J_\nu \phi_\nu d\nu~~~,
\end{equation}
where $\phi_\nu$ is the line profile function for the ionized gas in the position determined by electron temperature, density, microturbulent velocity field, and gas velocity \citep{zhu22}. The observed line intensity at the frequency $\nu$ is the difference between the total and continuum intensities as
\begin{equation}
I_{\nu,\textrm{L}}=I_{\nu}-I_{\nu,\textrm{C}}~~~.
\end{equation}

In the iterative method used in the new model, the departure coefficients at different locations in H II regions from the center to the boundary are derived by solving the matrix equation mentioned above without effects of radiation fields in the beginning. In the second step, the mean intensities of incident radiation fields at corresponding grid points are calculated from the derived departure coefficients. In the third step, the departure coefficients are computed with the calculated mean intensities of incident radiation fields. Then, this process is operated iteratively to derive convergent values of departure coefficients. The calculation continues until convergence is achieved at all grid points. The details of how to determine convergence in this model are the same as those given in the previous work \citep{zhu22}. In addition, the Lyman transitions are assumed to be optical thick, and the collisional excitations from the levels of $n=1$ and $n=2$ are neglected \citep{hum87,sto95,zhu22}. The new model in this work is improved from the model given in \citet{zhu22}. The main improvement of the new model is the non-local calculation of the incident radiation fields.

\subsection{Simplifications in the radiative transfer treatment of the new model} \label{sec:simple}

\begin{figure}
\begin{center}
\includegraphics[scale=0.5]{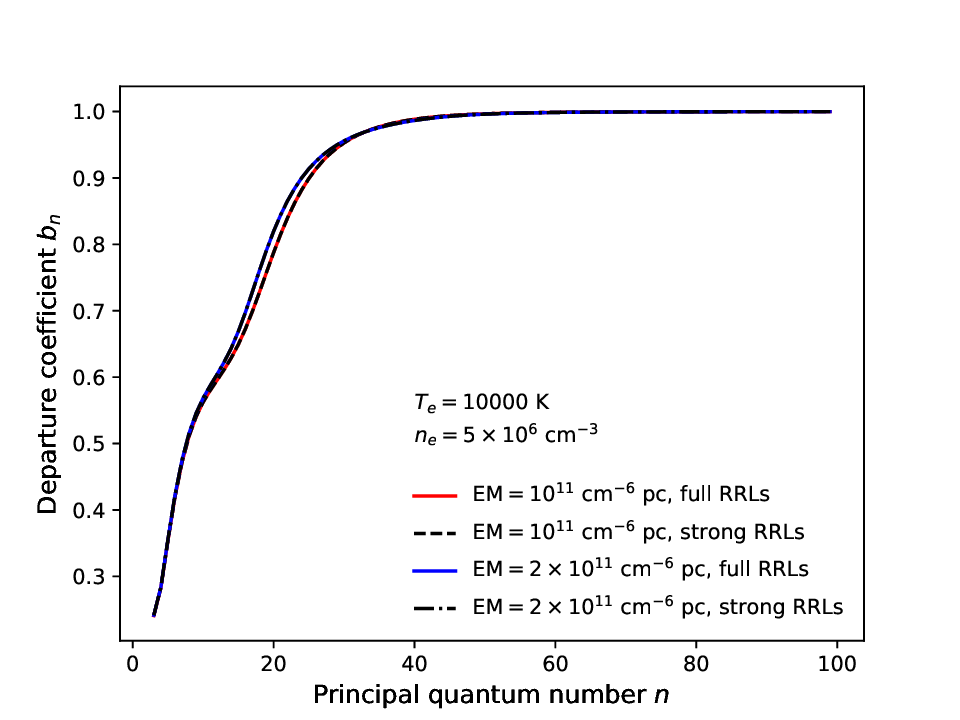}
\includegraphics[scale=0.5]{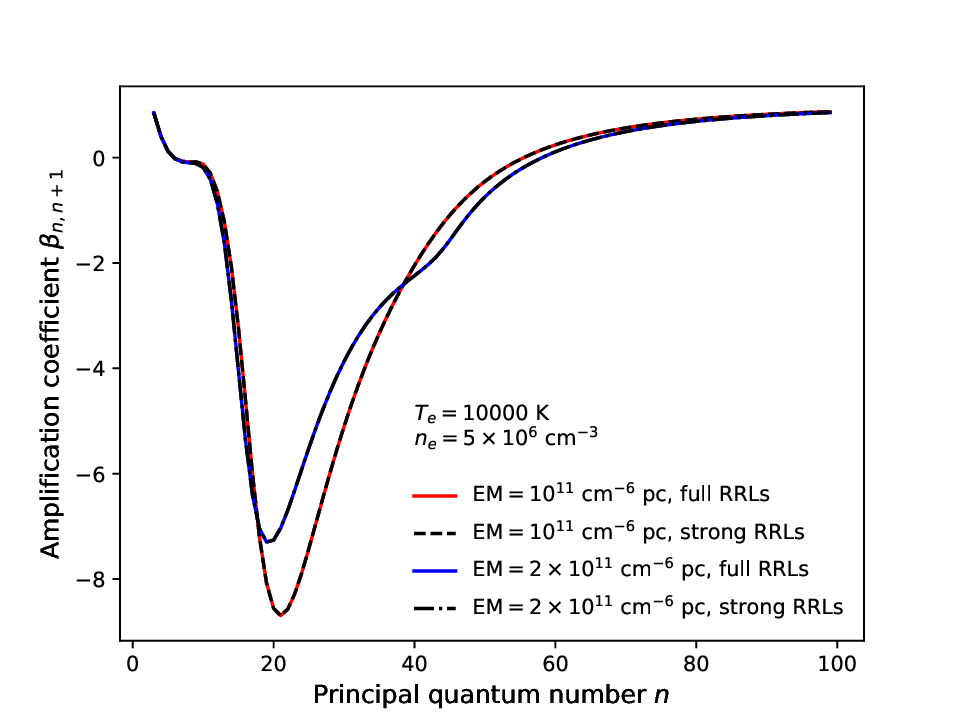}
\caption{The results calculated from the EPA models considering all hydrogen RRLs in the treatment of incident radiation fields are compared with those considering only relatively strong hydrogen RRLs. In the top panel, the $b_n$ in different cases are plotted. The $\beta_{n,n+1}$ is presented in the bottom panel.}\label{fig:simple_rt}
\end{center}
\end{figure}

In order to improve the computation speed of the new model, several simplifications are made in the treatment of radiative transfer. First, only the hydrogen RRLs with the difference between the upper and lower levels $\Delta n=m-n\leq10$ or with the lower level $n\leq10$ are calculated. This is easy to understand because the RRLs with $\Delta n>10$ are always very weak so that they have no effects on the level populations. The spontaneous transition probability $A_{m,n}$ first decreases and then increases when $n$ decreases from $m$-1 to 1 if $m\gg 1 $ \citep{bro71}. So the RRLs with lower level $n\leq10$ are included in calculations although it turned out that taking these lines into account has a very insignificant effect on the results. The results with/without this simplification in the EPA model are compared and shown in Fig. \ref{fig:simple_rt}, showing negligible differences. As verifying this simplification's reliability in the new model requires excessive computation, the EPA model is used. Although the two models differ in radiative transfer treatment, the effects of simplification are similar. Thus, it is appropriate to apply this simplification to the new model.

Second, when treating the radiative transfer of hydrogen RRLs, departure coefficients $b_n$ with principal quantum number $n$ and the spontaneous transition probability $A_{m,n}$ instead of $b_{nl}$ and $A_{ml',nl}$ are used. The value of $b_n$ is derived from 
\begin{equation}
b_n=\frac{1}{n^2}\sum^{n-1}_{l=0}(2l+1)b_{nl}~~~.
\end{equation}
The values of $b_{nl}$ are calculated from the level population equations. The average spontaneous transition probability $A_{m,n}$ from the states of principal quantum number $m$ to those of $n$ is
\begin{equation}
A_{mn}=\frac{1}{m^2}\sum_{l'}\sum_{l'=l\pm1}(2l'+1)A_{ml',nl}~~~.
\end{equation}
Then the Eqs. \ref{equ:cross} and \ref{equ:emission} can be simplified as
\begin{equation}
\kappa_{\nu,\textrm{L}}=\frac{h\nu}{4\pi}(N_nB_{nm}-N_mB_{mn})\phi_{\nu}=b_n\beta_{n,m}\kappa_{\nu,\textrm{L}}^{\textrm{LTE}}~~~,
\end{equation}
and
\begin{equation}
j_{\nu,\textrm{L}}=\frac{h\nu}{4\pi}\phi_\nu N_{m}A_{mn}=b_mB_\nu(T_e)\kappa_{\nu,\textrm{L}}^{\textrm{LTE}}~~~,
\end{equation}
where $B_{mn}$ can be derived from $A_{mn}$ \citep{pet12}. $\kappa_{\nu,\textrm{L}}^{\textrm{LTE}}$ is the line absorption coefficient under the LTE assumption. $\beta_{n,m}$ is the amplification coefficient with the upper and lower level $m$ and $n$ given by \citep{gor02,pet12}
\begin{equation}
\beta_{n,m}=\frac{1-\frac{b_m}{b_n}\exp(-\frac{h\nu}{kT})}{1-\exp(-\frac{h\nu}{kT})}~~~.
\end{equation}

\begin{figure}
\begin{center}
\includegraphics[scale=0.5]{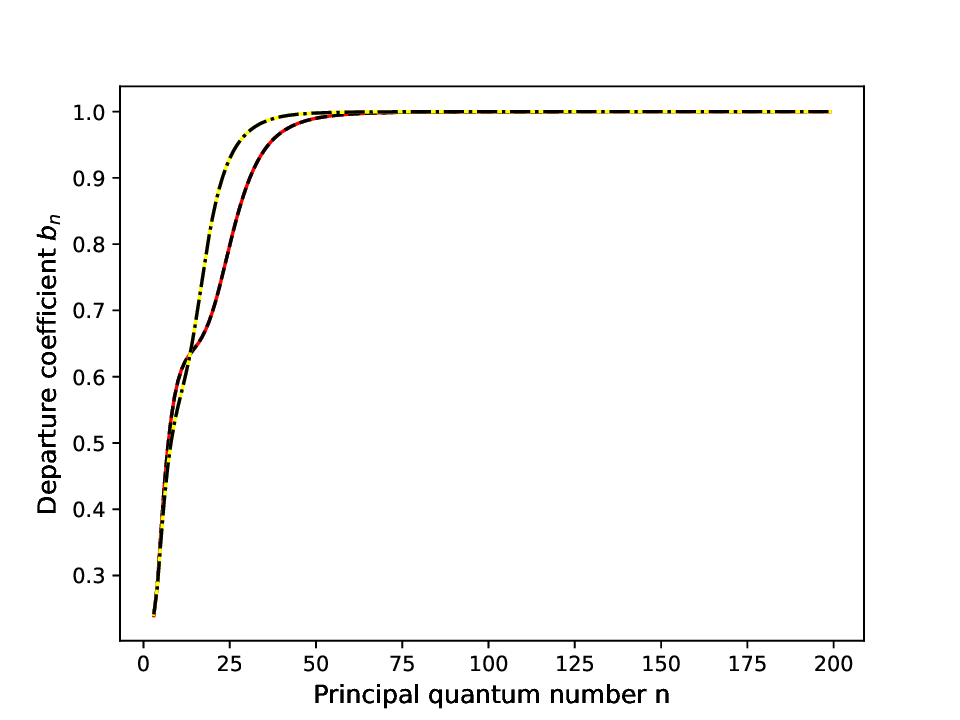}
\includegraphics[scale=0.5]{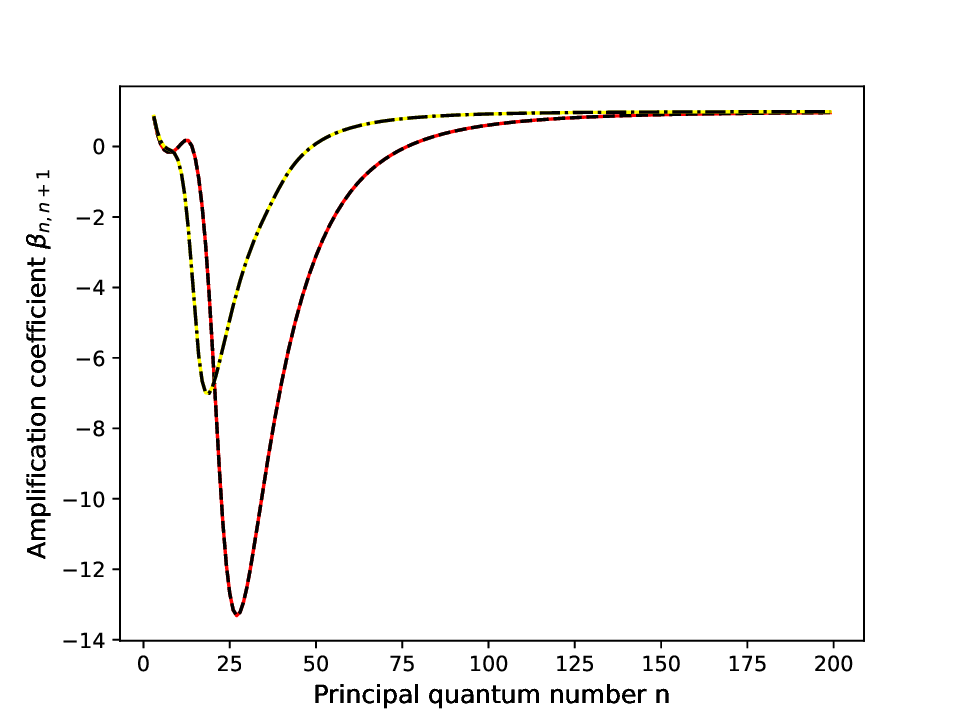}
\caption{The comparison of the results calculated from the improved new model with or without the second simplification mentioned in Section \ref{sec:simple}. The red and yellow solid lines are the results calculated without the simplification for the cases of $T_e=10000$ K, $n_e=10^6$ cm$^{-3}$, EM=$10^{10}$ cm$^{-6}$ pc and $T_e=10000$ K, $n_e=10^7$ cm$^{-3}$, EM=$10^{11}$ cm$^{-6}$ pc, respectively. The black dashed and dash-dotted lines are those calculated with the simplification for the corresponding cases. In the top panel, the $b_n$ at the center of uniform spherical H II regions are plotted. The $\beta_{n,n+1}$ is presented in the bottom panel.}\label{fig:simple_nl}
\end{center}
\end{figure}

In order to check the reliability of this simplification, the results derived from the simplified model are compared with those from the unsimplified model. As the examples shown in Fig. \ref{fig:simple_nl}, the differences due to this simplification are negligible.

\begin{figure}
\begin{center}
\includegraphics[scale=0.5]{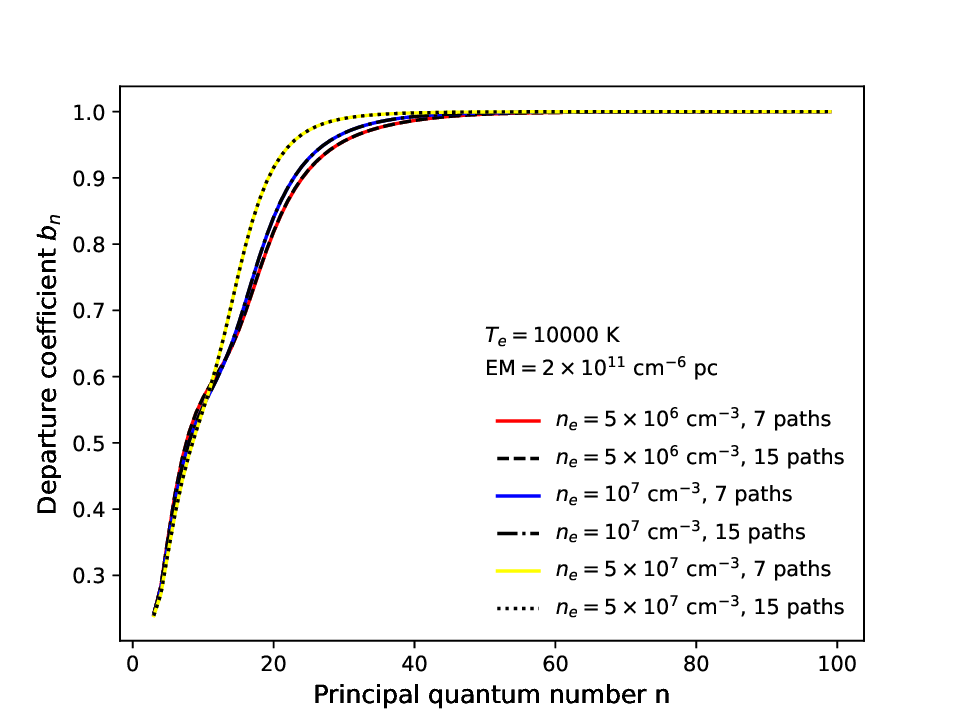}
\includegraphics[scale=0.5]{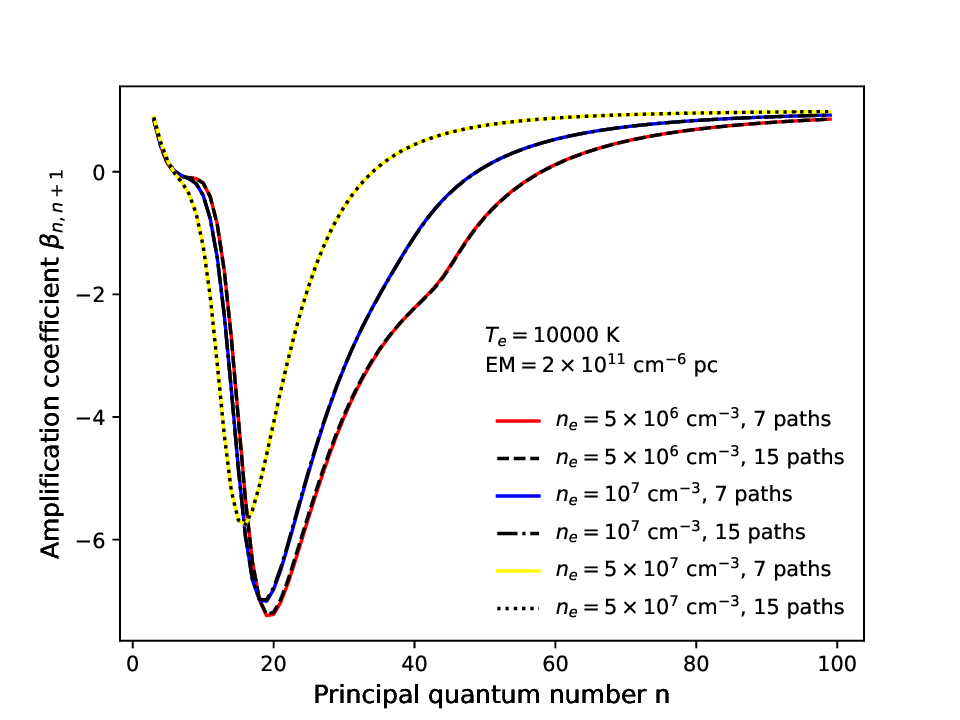}
\caption{The comparison of the results calculated from the improved new model with radiative transfer treated along 7 paths and 15 paths for some exemplary cases. In the top panel, the $b_n$ at the center of uniform spherical H II regions are plotted. The corresponding $\beta_{n,n+1}$ is presented in the bottom panel.}\label{fig:simple_drt}
\end{center}
\end{figure}

As shown in Fig. \ref{fig:transfer}, the radiative transfer of hydrogen RRLs and continuum is calculated along several paths in the two-dimensional axisymmetric coordinate system. Although a greater number of paths in radiative transfer calculations yields more accurate results, we typically restrict the computation to seven distinct paths at each spatial grid point requiring departure coefficient determination to optimize computational efficiency. By comparing computational results obtained with different numbers of paths, it is revealed that the inclusion of seven distinct paths in the radiation transfer calculations achieves sufficient accuracy while maintaining computational feasibility. This is shown in the comparison of the results derived from the calculations with 7 paths and 15 paths presented in Fig. \ref{fig:simple_drt}. In this work, when calculating departure coefficients with the improved model, the three aforementioned simplifications in radiative transfer are adopted.

In addition, although different numbers of the grid points from the center to the boundary of a spherical H II region can be set in the current model, the number of grid points is always 10 for the cases in this paper. The corresponding positions of these grid points are from $0.05R$ to $0.95R$ with equal spacing of $0.1R$ where $R$ is the radius of the spherical H II region. The model with 20 grid points is also applied to some exemplar cases. The results are not significantly different.

\subsection{The Instability of calculations with \textbf{extremely} high EM}

Emission measure (EM) is a measure of the ionized mass of a nebula emitting the free-free radiation \citep{gor02}. It is commonly defined as $EM=\int_{source} n_e^2 ds$ with the line-of-sight (LOS) path $s$. The EPA model becomes unstable when the EM is very high \citep{pro20}. This problem also appears in the current model with full radiative transfer treatment.  As the smoothing method used in \citet{pro20}, the values of intensity $J_\nu$ in adjacent steps are smoothed. The smoothed intensity of the incident radiation field $\bar{J_\nu}$ at the step $t$ is given by
\begin{equation}
\log(\bar{J_\nu})=\frac{1}{p}\sum^{t}_{k=t-p+1}\log(J_{\nu, k})~~~,
\end{equation}
where $J_{\nu, k}$ is the $J_\nu$ at the step $k$. $k$ means the sequence number of iterations. $p$ is the number of adjacent steps needed to be considered in smoothing. The value of $p$ is set to be $3-5$ in most cases, and to be 10 when the EM is extremely high for given $T_e$ and $n_e$. The smoothing method is useful to make calculations stable in some cases with high EMs. However, breakdown of the calculation could still happen when the EM exceeds a limit for given $T_e$ and $n_e$. Compared with the available observations \citep{woo89,pre20,bae13,san19}, the roughly estimated EMs corresponding to $T_e$ and $n_e$ in the hyper- and ultra-compact H II regions are within the limits of the current models.

\section{Results and discussions} \label{sec:result}

In this section, the results calculated using the models without effects of radiation fields \citep{zhu19}, with the escape probability approximation \citep{zhu22}, and with the full radiative transfer treatment are compared. The appropriate conditions of the previous models are investigated. The advantage of the improved current model is studied. The machine-learning model based on the current model and the EPA model is created.

The current model can handle spherically symmetric H II regions with density gradients and radial velocities. However, to enable comparison with the EPA model which is only applicable to homogeneous targets, all simulated cases in this work are uniform spherical H II regions. For these cases, the EM is defined to be $n_e^2D$ with the diameter of an H II region $D$. The Lyman continuum photon production rate for an ionizing source in an H II region is assumed to be lower than $1.5\times10^{50}$ s$^{-1}$ in this work. The value corresponds to the rate of a young massive star cluster \citep{ngu17,zhu22}. This allows the strengths of the simulated radiation fields in this work to be limited to an appropriate range. The ionized helium-to-hydrogen ratio $Y^+$ is always equal to 0.1.

\subsection{The modification of the escape probability in the EPA model}

\begin{figure*}
\begin{center}
\includegraphics[scale=0.5]{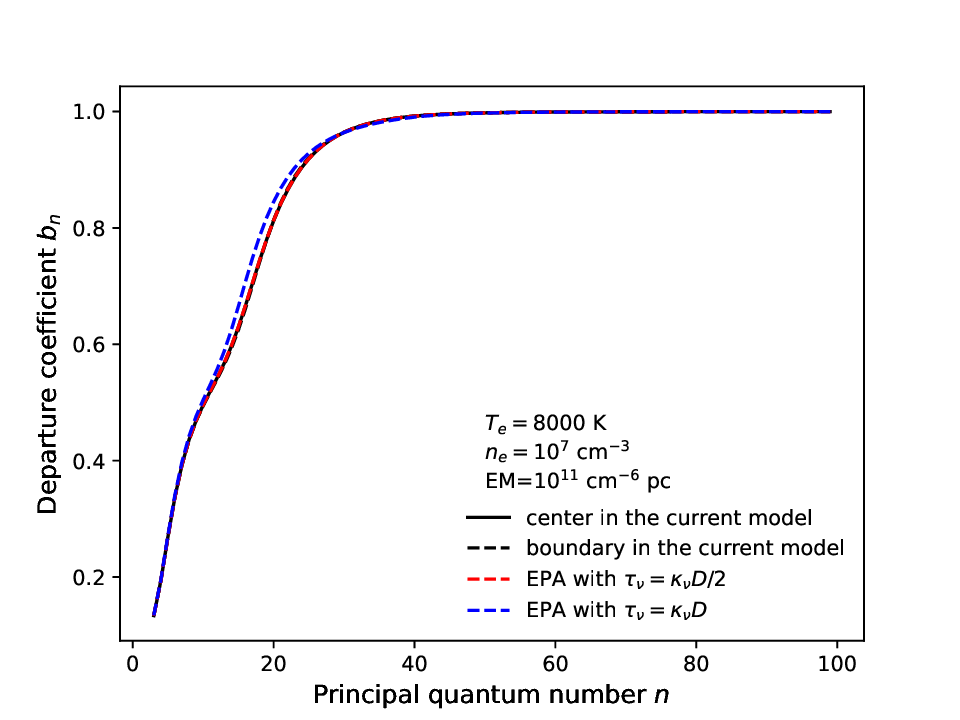}
\includegraphics[scale=0.5]{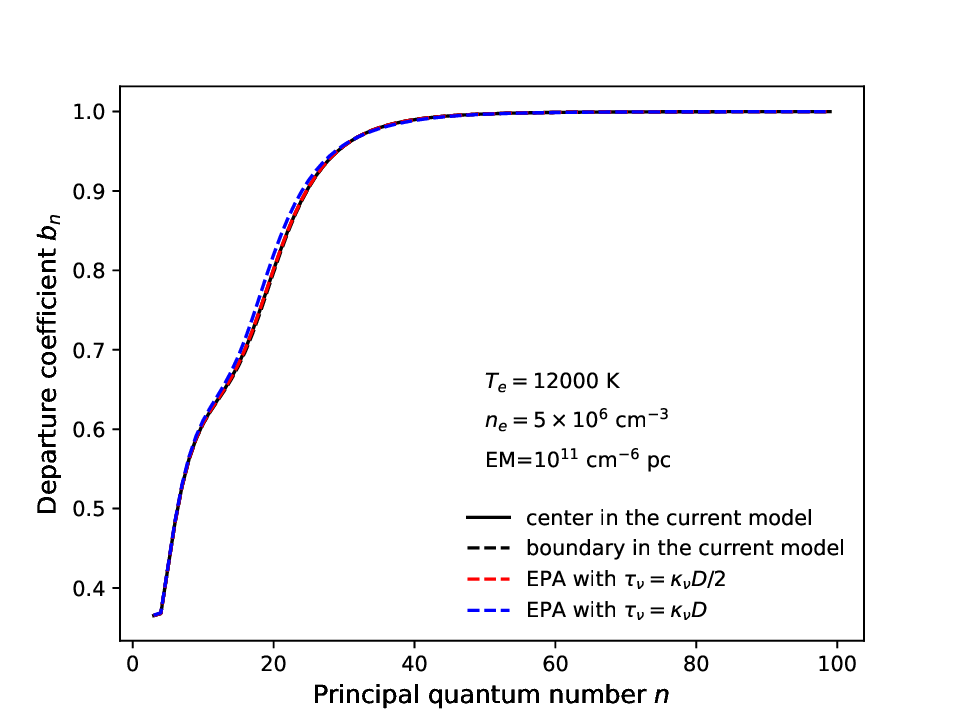}
\includegraphics[scale=0.5]{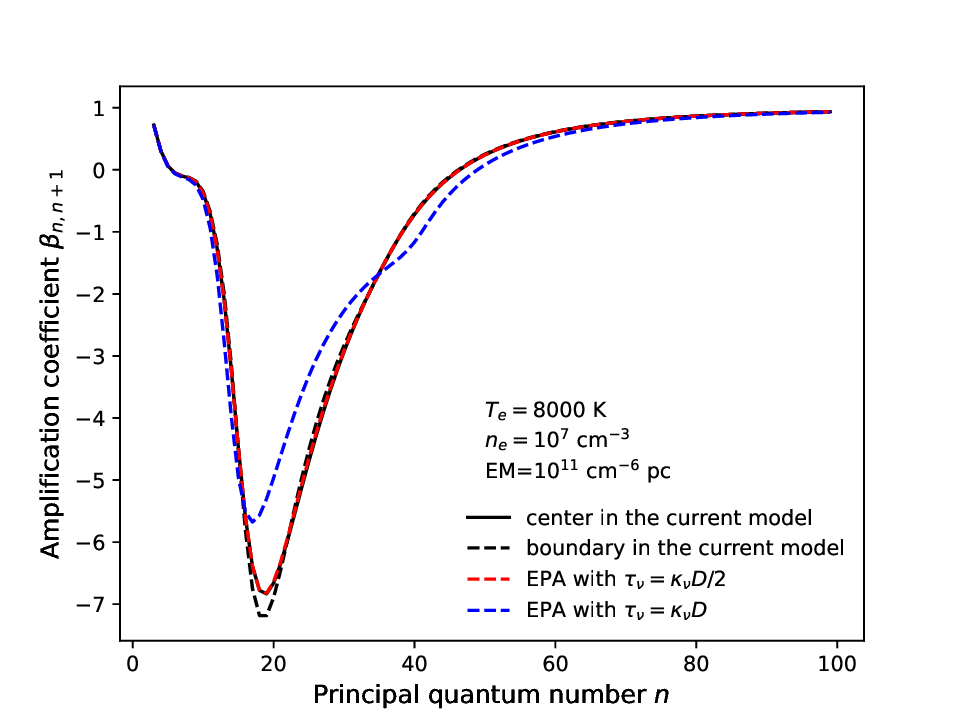}
\includegraphics[scale=0.5]{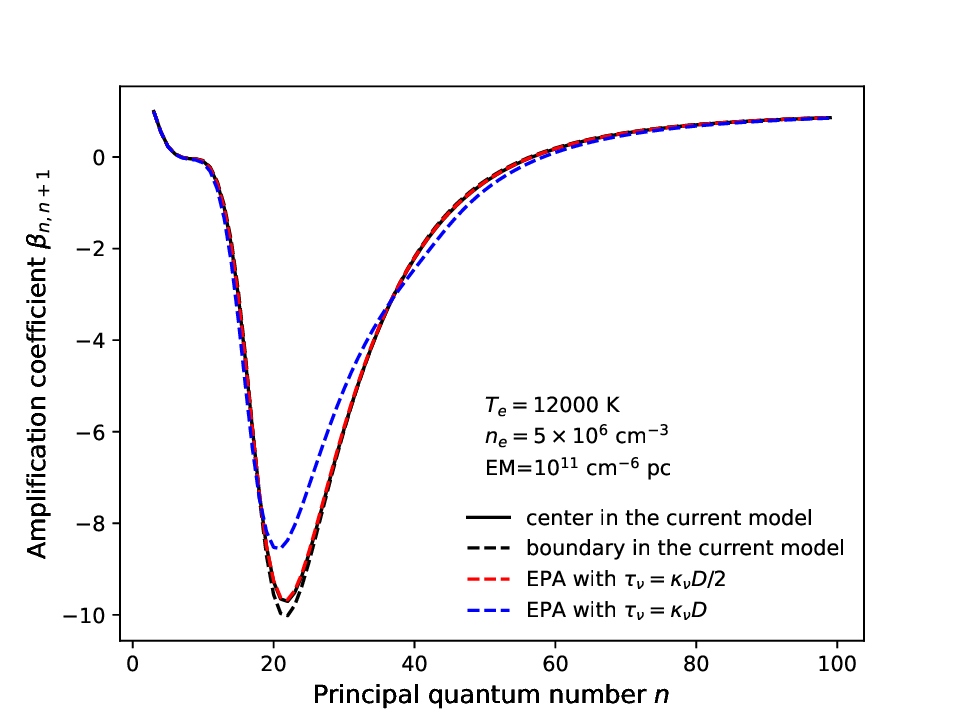}
\caption{The comparison of the results calculated from the improved current model with the full radiative transfer treatment and the EPA models with different assumptions of escape probability. In the top panel, the $b_n$ at the center and boundary of uniform spherical H II regions in the current model are compared with the EPA models with different assumptions of escape probability. The corresponding $\beta_{n,n+1}$ is presented in the bottom panel.}\label{fig:epatau}
\end{center}
\end{figure*}

The feasible escape probability of $\beta_\nu=e^{-\tau_\nu}$ is used in the EPA model given by \citet{pro20}. $\tau_\nu=\kappa_\nu L$ is the optical depth of a homogeneous medium of thickness $L$. For a uniform spherical H II region, the optical depth is $\tau_\nu=\kappa_\nu D$. Following \citet{pro20}, this assumption is also applied to the EPA model in our previous work \citep{zhu22}. However, from the comparison of the results calculated from the current improved model and the EPA model, it is found that $\tau_\nu=\kappa_\nu D/2$ is more appropriate as shown in Fig. \ref{fig:epatau}. It is easy to understand because the mean incident radiation field in the EPA model with $\tau_\nu=\kappa_\nu D/2$ is approximately equal to that at the center of the spherical H II region when the departure coefficients at different locations in the H II region are not significantly different. The escape probability of $\beta_\nu=e^{-\tau_\nu}$ with $\tau_\nu=\kappa_\nu D/2$ is consistent with those used in \citet{keg79} and \citet{koe80} for the CO and OH lines.

\subsection{Spatial variations in departure coefficients under different EM conditions within H II regions} \label{sec:EM}

When the radiation field is weak, the level populations of hydrogen atoms in H II regions are dominantly determined by electron temperature and density \citep{sto95,zhu22}. In this case, there is no significant difference between the departure coefficients in the inner and the boundary region. With increasing EM, the effects of radiation fields due to RRLs and continuum emission are gradually important. The spatial variations in departure coefficients within H II regions become more and more significant. In the top and middle panels of Fig. \ref{fig:departure_EM}, the departure coefficients $b_n$ calculated using different models are presented for the cases with different EMs. The differences between $b_n$ derived using different models increase with the value of EM. The results in the Case B model are significantly different from those in the other two models when EM is high. The differences between the $b_n$ in the center and boundary of H II region are presented in the bottom panel. It is presented that the spatial variations in departure coefficients within the H II region increase with EM except in few levels.

The corresponding comparisons for amplification coefficients $\beta_{n,n+1}$ are shown in Fig. \ref{fig:beta_EM}. The differences in the results between different models are clearer in the comparisons of $\beta_{n,n+1}$. The difference of the $\beta_{n,n+1}$ calculated between the EPA model and the current model is more significant in the case of EM$=6\times10^{11}$ cm$^{-6}$ pc than that in the case of EM$=2\times10^{11}$ cm$^{-6}$ pc. In the bottom panel, the difference between $\beta_{n,n+1}$ in the center and boundary of spherical H II region is displayed for the cases of EM$=2\times10^{11}$ cm$^{-6}$ pc and $6\times10^{11}$ cm$^{-6}$ pc. It demonstrates that $\beta_{n,n+1}$ and corresponding $b_n$ at different positions exhibit greater variation under the condition with a higher EM. In addition, it is presented in the top and middle panels that the effects of radiation fields on hydrogen level populations cause the population inversion to be weakened in the center and strengthened on the two sides of the valley of $\beta_{n,n+1}$. The relevant detailed explanation of this phenomenon is written in \citet{zhu22}.

The luminosities of Hn$\alpha$ lines are calculated using the three models for the two cases of different EMs but same $T_e$ and $n_e$.  The comparative differences between both the Case B and EPA models relative to the current model are illustrated in Fig. \ref{fig:flux_EM}. In the two cases, the luminosities simulated using the Case B model are both much different from those calculated using the other models. The differences of the Hn$\alpha$ line luminosities between the EPA model and the current model are all smaller than 10\% in the case of EM$=2\times10^{11}$ cm$^{-6}$ pc. While the differences become more significant in the case of EM$=6\times10^{11}$ cm$^{-6}$ pc.

From the comparison of the two cases with different EMs and identical temperature and density, it is revealed that the spatial variations in departure coefficients within H II regions increase with the EM. Although only the cases of $T_e=10000$ K and $n_e=10^7$ cm$^{-3}$ are presented in this work, the trend persists under other conditions. This is easy to understand. Different from the continuum emission, the intensity of stimulated emission depends heavily on the propagation path when the optical depth $\tau<-1$ due to high EM. Consequently, the incident radiation field is more different between the center and boundary of an H II region with higher EM.

\subsection{Spatial variations in $b_n$ under different $n_e$ conditions} \label{sec:density}

The amplification coefficients $\beta_{n,n+1}$ with principal quantum number $n$ calculated using the three models are plotted in Fig. \ref{fig:beta_EM_lowdensity} for the case of $T_e=10000$ K, $n_e=5\times10^6$ cm$^{-3}$, and EM$=2\times10^{11}$ cm$^{-6}$ pc. Compared to the results in the case of identical $T_e$ and EM but $n_e=10^7$ cm$^{-3}$ shown in the top panel of Fig. \ref{fig:beta_EM}, the discrepancies in $\beta_{n,n+1}$ between the EPA and current models are more pronounced in the case of lower electron density. It is also presented in Fig. \ref{fig:flux_EM_lowdensity} that the differences of the Hn$\alpha$ line luminosities between the EPA and current models in the case of $n_e=5\times10^6$ cm$^{-3}$ are more distinct than those in the case of $n_e=10^7$ cm$^{-3}$ mentioned above. It is because the absolute values of line optical depths $\tau_{\nu, \textrm{L}}=b_n\beta_{n,n+1}\tau_{\nu, \textrm{L}}^{\textrm{LTE}}$ of the masing Hn$\alpha$ lines are higher in the case of lower $n_e$ so that the non-isotropic stimulated emission is more powerful. 

\subsection{Spatial variations in $b_n$ under different $T_e$ conditions} \label{sec:temperature}

A similar trend also appears in the cases of low temperature. The results in the case of $T_e=6000$ K, $n_e=10^7$ cm$^{-3}$, and EM$=10^{11}$ cm$^{-6}$ pc are shown in Fig. \ref{fig:beta_EM_lowtemperature}. Compared with the cases of $T_e=10000$ K and $n_e=10^7$ cm$^{-3}$, the absorption coefficients $\kappa_{\nu, \textrm{C}}$ and $\kappa_{\nu, \textrm{L}}^{\textrm{LTE}}$ are higher in the cases of low $T_e$. This leads to higher absolute values of $\tau_{\nu}$ for masing lines. As shown in the top panel of Fig. \ref{fig:beta_EM_lowtemperature}, although the valley of $\beta_{n,n+1}$ is shallower in the case of $T_e=6000$ K, the higher $|\tau_{\nu}|$ for masing lines cause the stimulated emission to be powerful enough to make the amplification coefficients in the center of H II region significantly different from those in the boundary of H II region even when EM is $10^{11}$ cm$^{-6}$ pc. 

The differences of line luminosities between different models are presented in the bottom panel of Fig. \ref{fig:beta_EM_lowtemperature}. The difference between the EPA model and the current model is clearly higher than that in the case of $T_e=10000$ K, $n_e=10^7$ cm$^{-3}$, and EM=$2\times10^{11}$ cm$^{-6}$ pc for most millimeter Hn$\alpha$ lines.

\subsection{The effects of microturbulent velocity field}

The line profile function of a hydrogen RRL is affected by the microturbulent velocity field \citep{pet12}. In the previous work \citep{zhu22}, it was found that the importance of stimulated emission decreases with the increasing root mean square $\sigma_v$ of microturbulent velocity field. Consequently, the differences between the results in the EPA and current models decrease with the increasing $\sigma_v$. This is corroborated by the results shown in Fig. \ref{fig:beta_sigmav} in which the amplification coefficients $\beta_{n,n+1}$ and Hn$\alpha$ line luminosities in the case of $T_e=10000$ K, $n_e=10^7$ cm$^{-3}$, EM$=6\times10^{11}$ cm$^{-6}$ pc, and the root mean square $\sigma_v=0$ km s$^{-1}$ and 20 km s$^{-1}$ are displayed. The difference between the results calculated in the EPA model and the current model is lower in the case of higher $\sigma_v$. However, as shown in the bottom panel of Fig. \ref{fig:beta_sigmav}, the difference of line intensities is still significant even in the case of $\sigma_v=20$ km s$^{-1}$ under the condition of strong stimulated emission.

\subsection{The appropriate ranges of the previous models} \label{sec:epa_range}

In this section, the appropriate ranges of $T_e$, $n_e$, and EM for the Case B model and the EPA model are investigated in the cases of uniform spherical UC and HC H II regions. The classification standard for H II regions listed in \citet{fue20} is adopted to classify UC H II regions ($n_e\geq10^4$ cm$^{-3}$ and EM$\geq10^7$ cm$^{-6}$ pc) and HC H II regions ($n_e\geq10^6$ cm$^{-3}$ and EM$\geq10^{10}$ cm$^{-6}$ pc). 

\subsubsection{The Case B model}

The differences between the luminosities of the H20$\alpha$, H25$\alpha$, H30$\alpha$, H35$\alpha$, and H40$\alpha$ lines derived using the current model and the Case B model for UC and HC H II regions are shown in Tables \ref{table:deviation_low} and \ref{table:deviation_high_caseb}, respectively. These millimeter hydrogen RRLs are used to present the differences between different models because hydrogen RRL masers that can significantly affect the hydrogen level populations were detected in millimeter wavelengths \citep{mar89}. As mentioned above, the Lyman continuum photon production rate for an ionizing source in an H II region is assumed to be lower than $1.5\times10^{50}$ s$^{-1}$. The upper limit of EM is derived with given $T_e$ and $n_e$. Then the cases with EM higher than the upper limit are not calculated.

It is shown in Table \ref{table:deviation_low} that the differences between the line luminosities using the two models are smaller than $10\%$ both for the cases of UC H II regions with $T_e=6000$ K and $10000$ K. If a difference lower than $10\%$ is thought to be insignificant, the Case B model should be accurate enough for calculating the hydrogen RRLs in UC H II regions. The differences between the two models for HC H II regions are presented in Table \ref{table:deviation_high_caseb}. The Case B model is only appropriate to be applied to a small fraction of cases with relatively low EM $\sim10^{10}$ cm$^{-6}$ pc for HC H II regions.

In \citet{pro18}, it was implied that the Case B model is not suitable for UC H II regions because the effect of free-free continuum radiation on hydrogen level populations becomes significant for $n_e>10^4$ cm$^{-3}$. However, our results shown in Table \ref{table:deviation_low} and Table \ref{table:deviation_high_caseb} indicate that the Case B model is suitable for simulating hydrogen RRLs in UC H II regions, but probably exhibits substantial deviations for HC H II regions.

\subsubsection{The EPA model}

The differences between the line luminosities calculated using the current model and the EPA model for HC HII regions are written in Table \ref{table:deviation_high}. As mentioned in Sections \ref{sec:EM}, \ref{sec:density}, and \ref{sec:temperature}, the differences increase with EM, and decrease with $T_e$ and $n_e$. The conditions under which the results calculated by the two models are significantly different are shown. Although only the results for a few cases are listed, calculations for the cases with other values of $T_e$, $n_e$, and EM have also been done. According to the results, the EPA model is not accurate enough in the cases of EM > EM$_\textrm{crt}$ for given $T_e$ and $n_e$. The value of EM$_\textrm{crt}$ can be roughly estimated by an empirical formula as 
\begin{equation}
\log(EM_\textrm{crt}) \gtrsim 0.77\log(n_e)-\frac{1.74T_e}{10^5}(\log(n_e)-13.74)+4.92 ~~~,
\end{equation}
where $T_e$, $n_e$, and EM are in units of K, cm$^{-3}$, and cm$^{-6}$ pc, respectively.

Compared with the estimated values of $T_e$, $n_e$, and EM in previous observations toward HC H II regions, it seems to suggest that the EPA model is appropriate for most HC H II regions if these H II regions can be treated as a uniform and spherical ones \citep{mur10,pre20}. However, for rare objects with clearly detected hydrogen recombination line masers (e.g., MWC349 and MWC922), the EPA model becomes inadequate due to their extremely high EMs \citep{bae13,san19,pre20,mar23}, even without accounting for density gradients in these sources.

\begin{table*}\tiny
\centering
\caption{The differences between the luminosities of the Hn$\alpha$ lines calculated using the current model and the Case B model for a uniform spherical ultra-compact H II region with various $T_e$, $n_e$, and EM.} \label{table:deviation_low}
\begin{tabular}{|c|ccc|ccc|ccc|ccc|}
\hline
$T_e=6000$ K & \multicolumn{3}{c|}{$n_e=10^4$ cm$^{-3}$} & \multicolumn{3}{c|}{$n_e=5\times10^4$ cm$^{-3}$} & \multicolumn{3}{c|}{$n_e=10^5$ cm$^{-3}$} & \multicolumn{3}{c|}{$n_e=5\times10^5$ cm$^{-3}$} \\
EM [cm$^{-6}$ pc] & H30$\alpha$ & H35$\alpha$ & H40$\alpha$ &  H30$\alpha$ & H35$\alpha$ & H40$\alpha$ & H30$\alpha$ & H35$\alpha$ & H40$\alpha$ & H30$\alpha$ & H35$\alpha$ & H40$\alpha$ \\
\hline
$2\times10^7$ & 0.23\% & 0.23\% & 0.21\% & 0.17\% & 0.15\% & 0.12\% & 0.14\% & 0.11\% & 0.08\% & 0.07\% & 0.05\% & 0.03\% \\
$10^8$ & 0.16\% & 0.09\% & 0.01\% & 0.12\% & 0.10\% & 0.09\% & 0.11\% & 0.09\% & 0.06\% & 0.07\% & 0.04\% & 0.02\% \\
$2\times10^8$ & ... & ... & ... & 0.05\% & 0.04\% & 0.09\% & 0.06\% & 0.08\% & 0.06\% & 0.06\% & 0.04\% & 0.01\% \\
$10^9$ & ... & ... & ... & -1.03\% & -0.91\% & 2.04\% & -0.48\% & 0.36\% & 1.42\% & 0.11\% & 0.1\% & -0.05\% \\
$2\times10^9$ & ... & ... & ... & ... & ... & ... & ... & ... & ... & 0.48\% & 0.60\% & 0.10\% \\
\hline
$T_e=10000$ K & \multicolumn{3}{c|}{$n_e=10^4$ cm$^{-3}$} & \multicolumn{3}{c|}{$n_e=5\times10^4$ cm$^{-3}$} & \multicolumn{3}{c|}{$n_e=10^5$ cm$^{-3}$} & \multicolumn{3}{c|}{$n_e=5\times10^5$ cm$^{-3}$} \\
EM [cm$^{-6}$ pc] & H30$\alpha$ & H35$\alpha$ & H40$\alpha$ &  H30$\alpha$ & H35$\alpha$ & H40$\alpha$ & H30$\alpha$ & H35$\alpha$ & H40$\alpha$ & H30$\alpha$ & H35$\alpha$ & H40$\alpha$ \\
\hline
$2\times10^7$ & 0.22\% & 0.22\% & 0.21\% & 0.16\% & 0.15\% & 0.12\% & 0.13\% & 0.11\% & 0.08\% & 0.07\% & 0.05\% & 0.03\% \\
$10^8$ & 0.19\% & 0.16\% & 0.12\% & 0.14\% & 0.12\% & 0.09\% & 0.12\% & 0.10\% & 0.07\% & 0.07\% & 0.04\% & 0.02\% \\
$2\times10^8$ & ... & ... & ... & 0.11\% & 0.09\% & 0.07\% & 0.10\% & 0.08\% & 0.05\% & 0.06\% & 0.04\% & 0.02\% \\
$10^9$ & ... & ... & ... & -0.15\% & -0.16\% & 0.19\% & -0.05\% & 0.04\% & 0.14\% & 0.05\% & 0.03\% & -0.02\% \\
$2\times10^9$ & ... & ... & ... & ... & ... & ... & ... & ... & ... & 0.06\% & 0.05\% & -0.02\% \\
\hline
\end{tabular}
\end{table*}

\begin{table*}\tiny
\centering
\caption{The differences between luminosities of the Hn$\alpha$ lines calculated using the current model and the Case B model for a uniform spherical hyper-compact H II region with various $T_e$, $n_e$, and EM.} \label{table:deviation_high_caseb}
\begin{tabular}{|c|ccc|ccc|ccc|}
\hline
$T_e=6000$ K & \multicolumn{3}{c|}{$n_e=10^6$ cm$^{-3}$} & \multicolumn{3}{c|}{$n_e=10^7$ cm$^{-3}$} & \multicolumn{3}{c|}{$n_e=10^8$ cm$^{-3}$} \\
EM [cm$^{-6}$ pc] & H30$\alpha$ & H35$\alpha$ & H40$\alpha$ &  H25$\alpha$ & H30$\alpha$ & H35$\alpha$ & H20$\alpha$ & H25$\alpha$ & H30$\alpha$ \\
\hline
$2\times10^{10}$ & $>$500\% & $>$500\% & 439.93\% &  1.74\% & -2.04\% & -2.68\% & -0.06\% & -0.15\% & -0.10\% \\
$10^{11}$ & ... & ... & ... & $>$500\% & $>$500\% & -71.22\% & -2.21\% & -5.63\% & -2.47\% \\
$2\times10^{11}$ & ... & ... & ... & $>$500\% & $>$500\% & 139.90\% & 16.1\% & -35.64\% & -11.62\% \\
$10^{12}$ & ... & ... & ... & ... & ... & ... & $>$500\% & -71.95\% & -99.39\% \\
\hline
$T_e=10000$ K & \multicolumn{3}{c|}{$n_e=10^6$ cm$^{-3}$} & \multicolumn{3}{c|}{$n_e=10^7$ cm$^{-3}$} & \multicolumn{3}{c|}{$n_e=10^8$ cm$^{-3}$} \\
EM [cm$^{-6}$ pc] & H30$\alpha$ & H35$\alpha$ & H40$\alpha$ &  H25$\alpha$ & H30$\alpha$ & H35$\alpha$ & H20$\alpha$ & H25$\alpha$ & H30$\alpha$ \\
\hline
$2\times10^{10}$ & 6.05\% & 6.33\% & 2.31\% &  0.07\% & -0.16\% & -0.29\% & 0.00\% & -0.02\% & -0.02\% \\
$10^{11}$ &        $>$500\% & $>$500\% & $>$500\% & 10.14\% & -2.36\% & -8.33\% & -0.13\% & -0.41\% & -0.37\% \\
$2\times10^{11}$ & ... & ... & ... & 160.58\% & 38.07\% & -39.57\% & -0.52\% & -1.84\% & -1.43\% \\
$10^{12}$ & ... & ... & ... & $>$500\% & $>$500\% & $>$500\% & 200.69\% & -61.80\% & -41.15\% \\
$2\times10^{12}$ & ... & ... & ... & ... & ... & ... & $>$500\% & -45.26\% & -94.52\% \\
\hline
\end{tabular}
\end{table*}

\begin{table*}\tiny
\centering
\caption{The differences between luminosities of the Hn$\alpha$ lines calculated using the current model and the EPA model for a uniform spherical hyper-compact H II region with various $T_e$, $n_e$, and EM.} \label{table:deviation_high}
\begin{tabular}{|c|ccc|ccc|ccc|}
\hline
$T_e=6000$ K & \multicolumn{3}{c|}{$n_e=10^6$ cm$^{-3}$} & \multicolumn{3}{c|}{$n_e=10^7$ cm$^{-3}$} & \multicolumn{3}{c|}{$n_e=10^8$ cm$^{-3}$} \\
EM [cm$^{-6}$ pc] & H30$\alpha$ & H35$\alpha$ & H40$\alpha$ &  H25$\alpha$ & H30$\alpha$ & H35$\alpha$ & H20$\alpha$ & H25$\alpha$ & H30$\alpha$ \\
\hline
$2\times10^{10}$ & 12.20\% & 30.99\% & 38.08\% &  -0.34\% & 0.04\% & 0.20\% & 0.00\% & 0.01\% & 0.00\% \\
$10^{11}$ & ... & ... & ... & 40.11\% & 57.35\% & 28.66\% & 0.14\% & 0.84\% & 0.46\% \\
$2\times10^{11}$ & ... & ... & ... & 104.63\% & 161.03\% & 165.89\% & 4.20\% & 0.57\% & 1.28\% \\
$10^{12}$ & ... & ... & ... & ... & ... & ... & 131.8\% & 193.61\% & 60.20\% \\
\hline
$T_e=10000$ K & \multicolumn{3}{c|}{$n_e=10^6$ cm$^{-3}$} & \multicolumn{3}{c|}{$n_e=10^7$ cm$^{-3}$} & \multicolumn{3}{c|}{$n_e=10^8$ cm$^{-3}$} \\
EM [cm$^{-6}$ pc] & H30$\alpha$ & H35$\alpha$ & H40$\alpha$ &  H25$\alpha$ & H30$\alpha$ & H35$\alpha$ & H20$\alpha$ & H25$\alpha$ & H30$\alpha$ \\
\hline
$2\times10^{10}$ & -1.19\% & -1.38\% & -1.33\% &  -0.06\% & -0.06\% & -0.06\% & 0.00\% & 0.00\% & -0.01\% \\
$10^{11}$ &             19.88\% & 42.08\% & 56.24\% & -0.78\% & -0.28\% & 0.28\% & -0.01\% & 0.04\% & 0.03\% \\
$2\times10^{11}$ & ... & ... & ... & 6.14\% & 7.81\% & -1.54\% & -0.01\% & 0.27\% & 0.25\% \\
$10^{12}$ & ... & ... & ... & 114.65\% & 199.59\% & 307.31\% & 15.42\% & 3.77\% & 1.49\% \\
$2\times10^{12}$ & ... & ... & ... & ... & ... & ... & 61.25\% & 76.77\% & -16.11\% \\
\hline
\end{tabular}
\end{table*}

\subsection{The advantage of the current model with full radiative transfer treatment}

In Section \ref{sec:epa_range}, the appropriate ranges of the Case B model and the EPA model are studied. The Case B model and the EPA model are suitable for ultra-compact H II regions and other classes of H II regions with lower $n_e$ and EM. The EPA model is appropriate to estimate the global properties of most cases of HC H II regions with EM$<$EM$_{\textrm{crt}}$. For HC H II regions with EM$\geq$EM$_{\textrm{crt}}$, the current model is necessary. Furthermore, in modeling the 2D observational images of the hydrogen RRLs from HC H II regions, the superiority of the current model over the EPA model becomes more pronounced, as the EPA model was primarily designed to simulate global-scale properties of an H II region \citep{pro20}.

\subsection{The machine-learning model for uniform spherical H II regions}

Calculations of departure coefficients of hydrogen level populations are often time-consuming. Especially for the current model with full radiative transfer, the calculation time can sometimes reach several days. This motivates us to train a machine-learning model to accelerate the calculation of departure coefficients. In this work, we employ the random forest algorithm, a powerful ensemble learning method widely used in various fields \citep{bre01}. It combines multiple decision trees to improve predictive performance. Each tree is built using a random subset of the training data and a random selection of features, reducing overfitting and correlation among trees. The final prediction is derived by aggregating the predictions of all trees.

Our training and validation data consist of the data simulated by using the current and EPA models. In a sample of the dataset, the values of $T_e$, log($n_e$), log(EM), and $\sigma_v$ are the inputs while $\beta_{n,n+1}$ with the principal quantum number $n$ in the range of 3 to 200 and $b_{200}$ are the outputs. The departure coefficients $b_n$ in the corresponding levels $n$ can be derived from the values of $\beta_{n,n+1}$ and $b_{200}$ with $T_e$. 

\subsubsection{Mock data and parameter settings for the machine-learning model}

In the dataset, the ranges of $T_e$, $n_e$, EM, and $\sigma_v$ are 5000-15000 K, $10^2-10^8$ cm$^{-3}$, $\leq10^{12}$ cm$^{-6}$ pc, and $\leq50$ km s$^{-1}$, respectively. The value of EM is also limited by the reasonable Lyman continuum photon production rate $<1.5\times10^{50}$ s$^{-1}$ for a spherical H II region with given $T_e$ and $n_e$. The value of $Y^+$ is assumed to be 0.1. According to the values of $T_e$, $n_e$, and EM, the total dataset is divided into three subsets as listed in Table \ref{table:dataset}. These three subsets include 12,168 samples with $n_e<10^4$ cm$^{-3}$, 401,335 samples with $n_e\geq10^4$ cm$^{-3}$ and EM$\lesssim0.5$EM$_\textrm{crt}$, and 39,292 samples of $n_e\geq10^4$ cm$^{-3}$ and EM$\geq0.5$EM$_\textrm{crt}$, respectively. The data in Subsets 1 and 2 are calculated by the EPA model, and those in Subset 3 are simulated by the current model with full radiative transfer treatment. Then, the total machine-learning model is composed of three sub-models corresponding to three subsets.  For each sub-model, the data in the subset is divided into a training set and a testing set with a ratio of 8 to 2. The former set is used to train the predictive model, and the latter one is used to assess the performance of that model. The function of random forest regressor\footnote{\url{https://scikit-learn.org/stable/modules/generated/sklearn.ensemble.RandomForestRegressor.html}} is used. The parameters set in this function are: the number of trees is 100; the maximum depth of the tree is set to be 13; the minimum number of samples required to split an internal node is 2; and the minimum number of samples required to be at a leaf node is 1.

\begin{table}\tiny
\centering
\caption{Classification of subsets in the total dataset. EM is also limited by the Lyman continuum photon production rate $<1.5\times10^{50}$ s$^{-1}$ for an H II region.} \label{table:dataset}
\begin{tabular}{|c|c|c|c|}
\hline
 & $T_e$ [K] & $n_e$ [cm$^{-3}$] & EM [cm$^{-6}$ pc] \\
 \hline
 Total dataset & 5000-15000 & $10^2-10^8$ & $\leq10^{12}$ \\
 \hline
 Subset 1 & 5000-15000 & $10^2-10^4$ & \\
 Subset 2 & 5000-15000 & $10^4-10^8$ & $\lesssim0.5$EM$_\textrm{crt}$ \\
 Subset 3 & 5000-15000 & $10^4-10^8$ & $\geq0.5$EM$_\textrm{crt}$ \\
\hline
\end{tabular}
\end{table}

The electron temperature, density-EM, and $\sigma_v$ distributions of the samples in the three subsets are shown in Fig. \ref{fig:tem_sample}. The temperature distributions are plotted in the top panels. The distribution for the samples of $n_e<10^4$ cm$^{-3}$ is relatively uniform. But the samples with high temperature are relatively fewer in the other two subsets because the effects of radiation fields are more important for the cases with low temperature so that more samples are needed for the machine-learning model to reach accurate predictions. The electron density-EM distributions are presented in the middle panels. The $n_e$-EM distributions in the samples are more inhomogeneous than the temperature distributions. For the cases of $n_e<10^4$ cm$^{-3}$, the effects of radiation fields due to continuum emission and RRLs on departure coefficients can be neglected as shown in Section \ref{sec:epa_range}. The samples are mainly distributed in EM$=20$ cm$^{-6}$ pc and $2.0\times10^7$ cm$^{-6}$ pc in the corresponding subset. The value of $b_n$ with EM between the two values can be correctly predicted since the different values of EM have neglectful effects on departure coefficients for the cases with EM<$10^{10}$ cm$^{-6}$ pc. In the second subset, a large part of the samples are distributed in the range of EM$>10^9$ cm$^{-6}$ pc. This is because departure coefficients are sensitive to the value of EM in the cases of EM close to EM$_{\textrm{crt}}$. More samples are necessary to be recorded for such cases. This kind of treatment is also shown in the $n_e-$EM distributions of the third subset plotted in the middle-right panel. The $\sigma_v$ distributions are presented in the bottom panels of Fig. \ref{fig:tem_sample}. The distribution for the first subset is not plotted. Since the effects of radiation fields are not significant in the cases of $n_e<10^4$ cm$^{-3}$, the values of $\sigma_v$ are all assumed to be 0 km s$^{-1}$.

\subsubsection{Validation results of machine-learning model}

\begin{figure*}
\begin{center}
\includegraphics[scale=0.35]{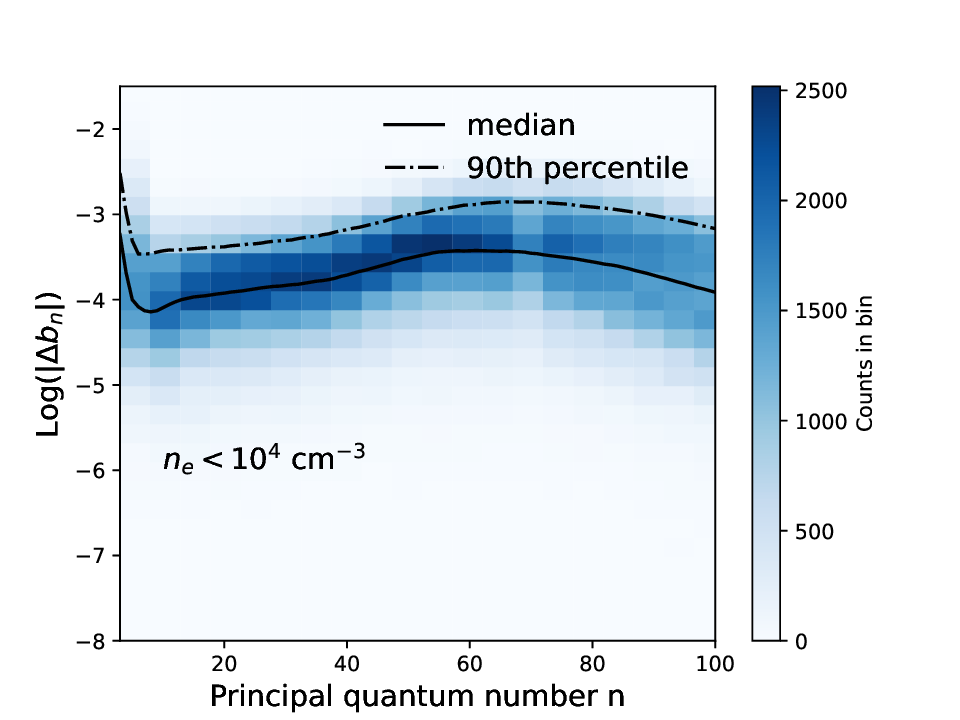}
\includegraphics[scale=0.35]{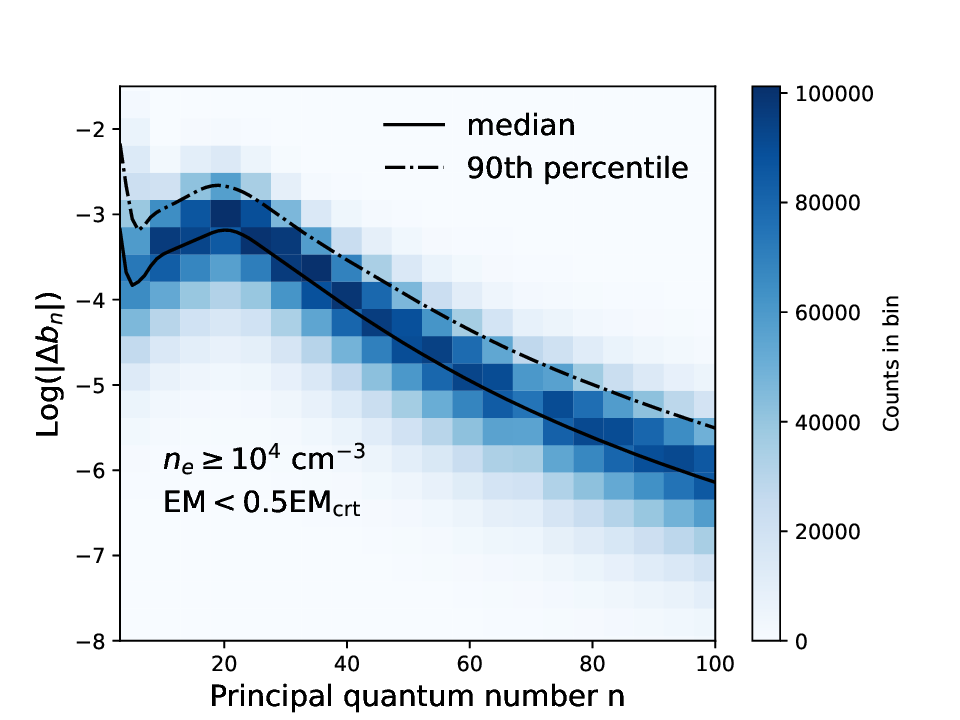}
\includegraphics[scale=0.35]{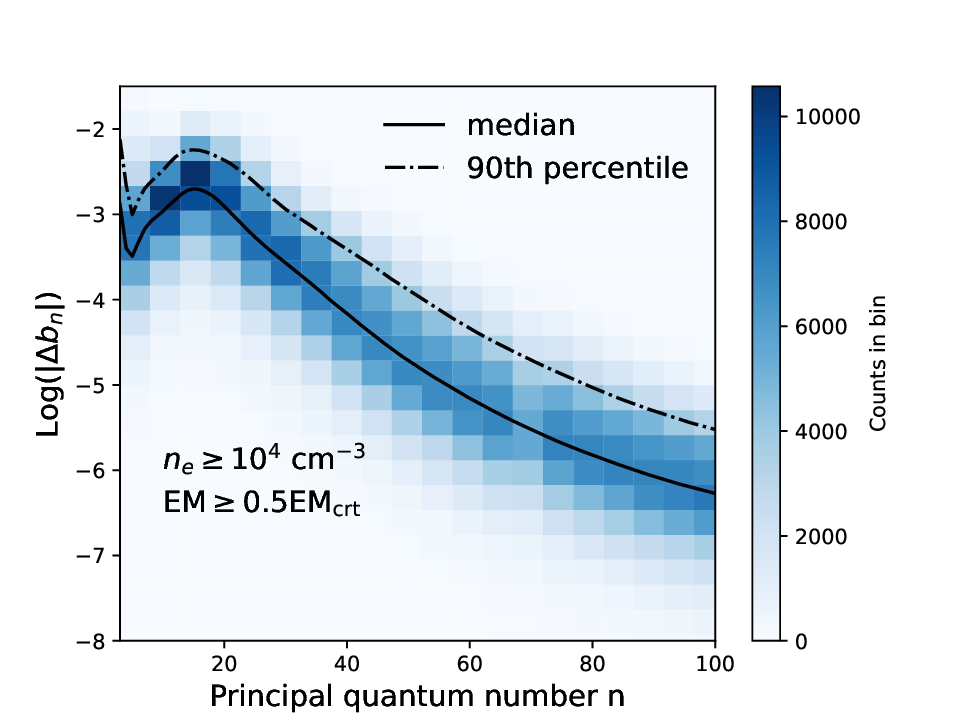}
\includegraphics[scale=0.35]{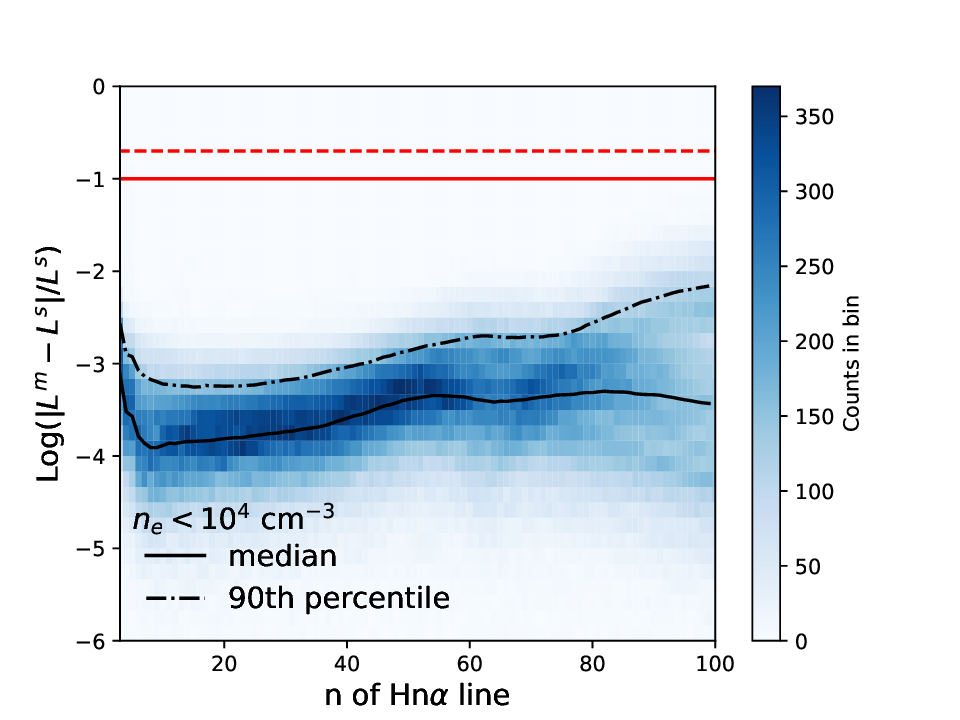}
\includegraphics[scale=0.35]{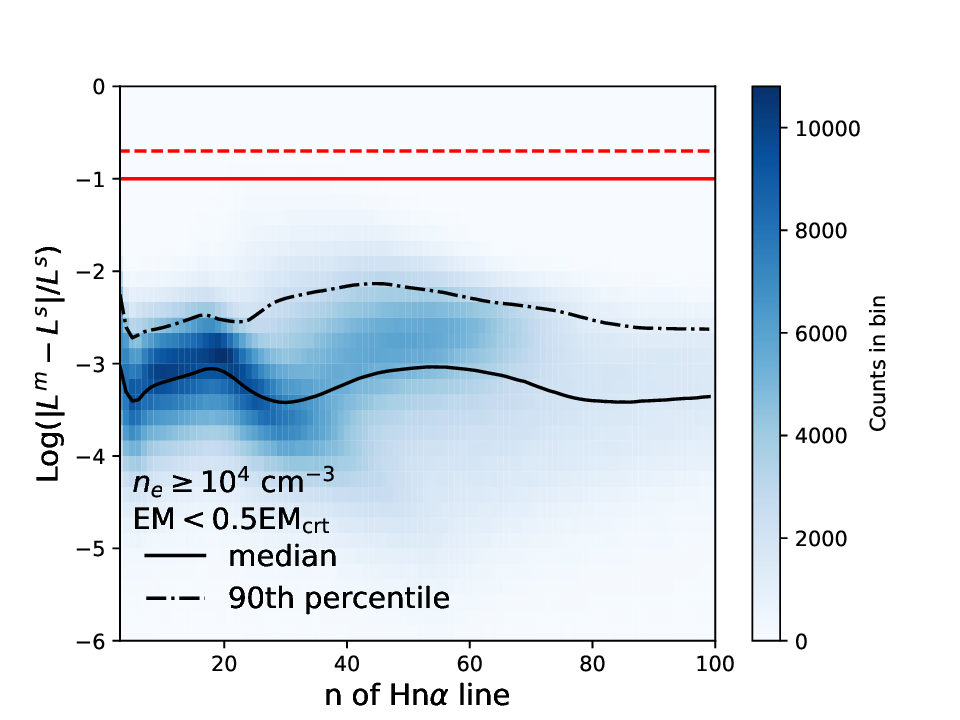}
\includegraphics[scale=0.35]{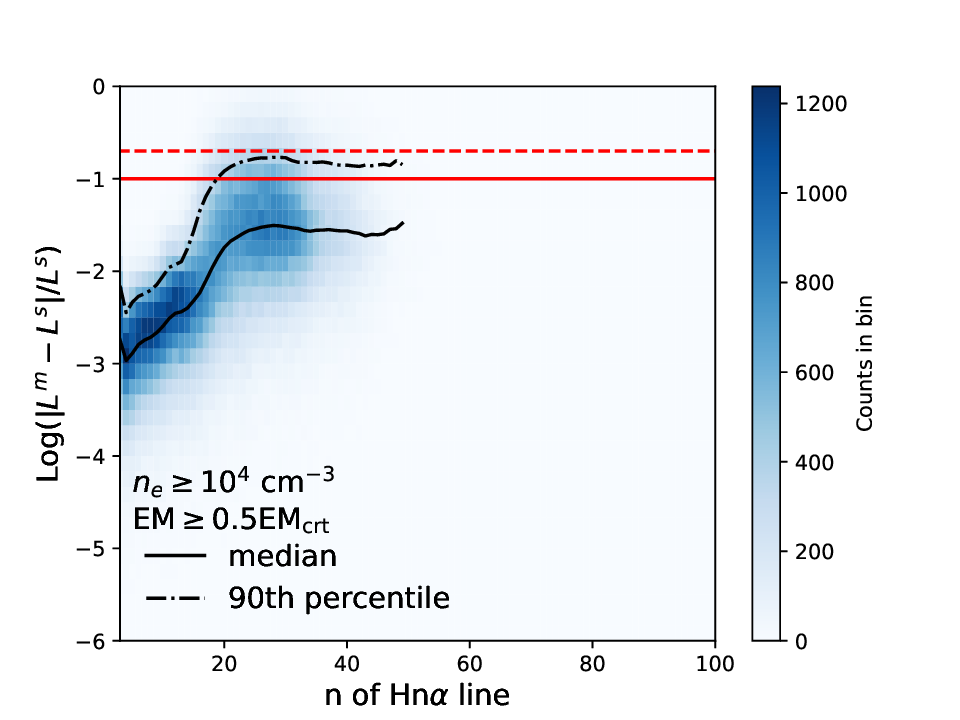}
\caption{The validation results of the machine-learning model on the testing set. The results of the three sub-models corresponding to different subsets are presented in the left, middle, and right panels, respectively. The comparisons of the departure coefficients calculated by the simulation models and the machine-learning sub-models are presented in the top panels. In the bottom panels, the comparisons of the Hn$\alpha$ line luminosities are shown. $L^m$ and $L^s$ denote the line luminosities computed using departure coefficients from the machine-learning model predictions and the original simulation models, respectively. The solid and dashed red lines in the bottom panels correspond to error levels of $10\%$ and $20\%$, respectively.}\label{fig:bn_sample}
\end{center}
\end{figure*}

The validation results of the testing sets are presented in Fig. \ref{fig:bn_sample}. The predicted and true values are very close. As presented in the top panels of Fig. \ref{fig:bn_sample}, the absolute differences between the departure coefficients $|\Delta b_n|$ calculated by the numerical models and predicted by the machine-learning model are mostly lower than 0.001. $b_n$ is very close to 1 at the high principal quantum number $n$ when the electron density $n_e$ is high \citep{bro70,sto95,pro18,zhu22}. This causes that the absolute errors of $b_n$ with high $n$ in the test sets of Subsets 2 and 3 are much smaller than those of Subset 1.

The Hn$\alpha$ line luminosities calculated using the $b_n$ from the simulation models and the machine-learning model are compared in the bottom panels of Fig. \ref{fig:bn_sample}. Cases with the continuum optical depth $\tau_{\nu, \textrm{C}}>3$ are not taken into account because practical observational strategies for hydrogen RRLs typically avoid such high $\tau_{\nu, \textrm{C}}$. So in the bottom-right panel for Subset 3, the results for Hn$\alpha$ lines with $n>50$ are not presented. For Subsets 1  and 2, the difference in the Hn$\alpha$ line luminosity is always lower than $1\%$. For Subset 3, the difference is lower than 10$\%$ in most cases although the difference could be higher in some conditions. This could result from the fewer samples in Subset 3 than those in Subset 2. In summary, the illustrations presented in Fig. \ref{fig:bn_sample} demonstrate that the machine-learning model achieves sufficient accuracy to substitute the numerical models for most spherical H II region cases. The machine-learning model and its datasets are available for download\footnote{\url{https://chemiverse.zero2x.org/models?model=HRRL}}.

\section{Summary and conclusions}\label{sec:conclusion}

In this work, improved from our previous EPA model, a new $nl$-model with the full radiative transfer treatment about hydrogen RRLs and continuum emission is developed for spherical H II regions. The departure coefficients in different locations within H II regions can be meticulously calculated.

The appropriate ranges of the Case B model and the EPA model are studied by comparing them with the new model. The effects of radiation fields on departure coefficients can be neglected for classical and UC H II regions ($n_e<10^6$ cm$^{-6}$ and EM$<10^9$ cm$^{-6}$ pc). With the appropriately modified escape probability, the EPA model is proven to be suitable for most cases of HC H II regions in which the EM is lower than the critical value of EM (EM$_\textrm{crt}$). By comparing the intensities of hydrogen RRLs calculated using the EPA model and the current model, the EM$_\textrm{crt}$ as a function of given $T_e$ and $n_e$ of a uniform and spherical H II region is estimated. Compared with our previous models, the new model is found to have a significant advantage for HC H II regions with extremely high EMs that are higher than EM$_\textrm{crt}$.

Based on a large sample of departure coefficients calculated by using the EPA model and the current model, an accurate machine-learning model is created to quickly estimate the departure coefficients and resulting properties of hydrogen RRLs from the electron temperature, number density, EM, and the root mean square of the microturbulent velocity field of a spherical H II region. The random forest algorithm is applied. The differences between the departure coefficients predicted by the machine-learning model and the simulation model are mostly lower than 0.001.

\begin{acknowledgements}
The authors thank the anonymous referee for the detailed comments to improve the quality of the manuscript. The work is supported by the National Natural Science Foundation of China No. 12373026, the Leading Innovation and Entrepreneurship Team of Zhejiang Province of China No. 2023R01008, the Key R\&D Program of Zhejiang, China No. 2024SSYS0012, and Zhejiang Provincial Natural Science Foundation of China No. LY24A030001.

\end{acknowledgements}

\clearpage

\appendix

\section{Additional figures}

\begin{figure}
\begin{center}
\includegraphics[scale=0.5]{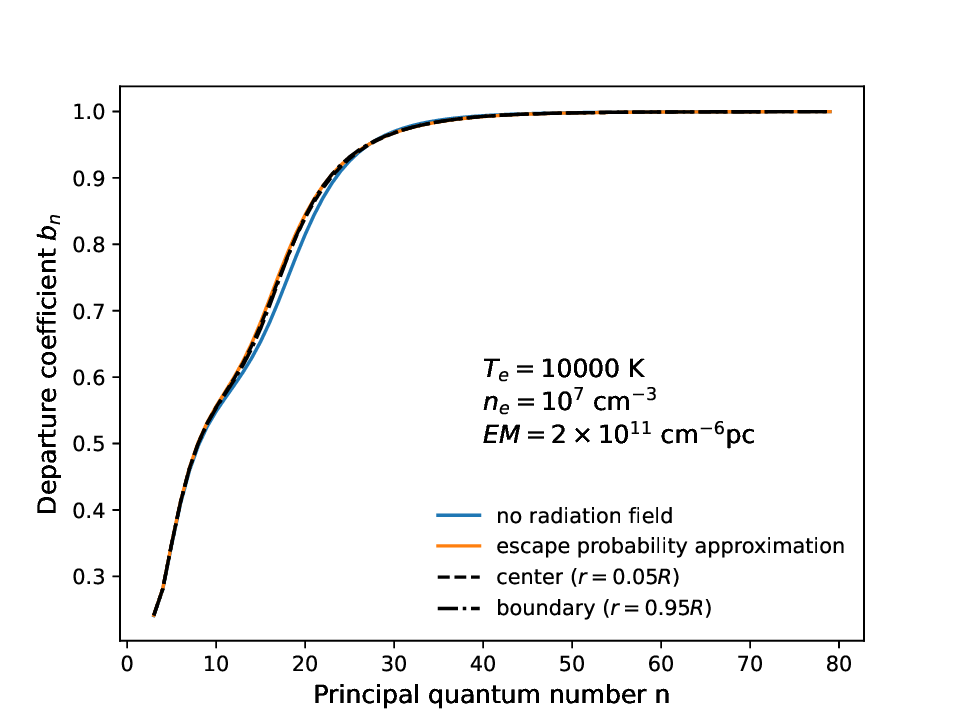}
\includegraphics[scale=0.5]{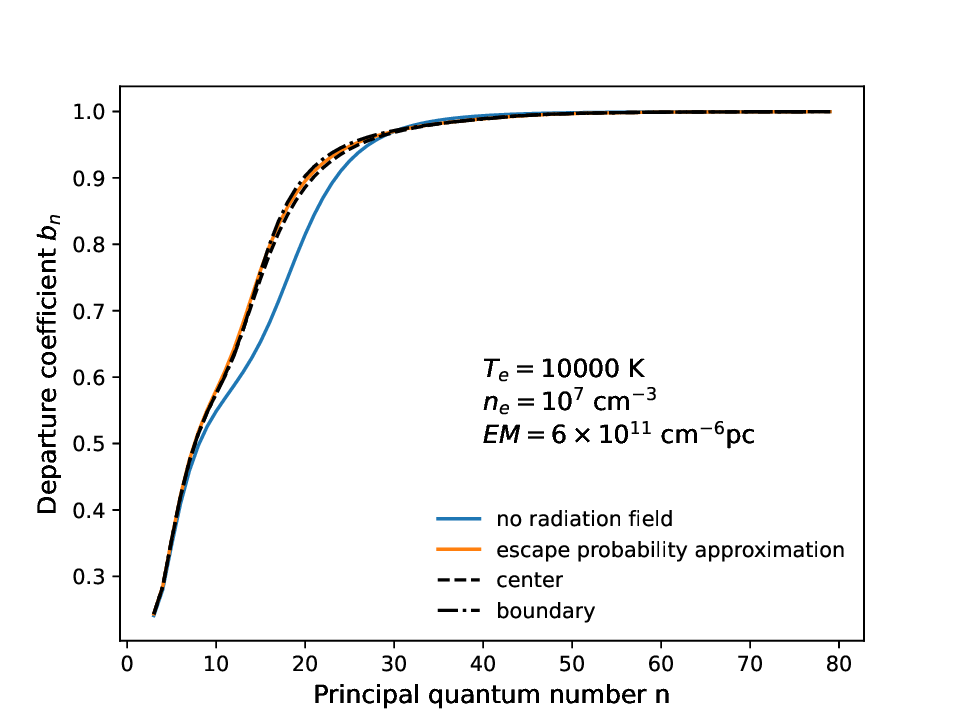}
\includegraphics[scale=0.5]{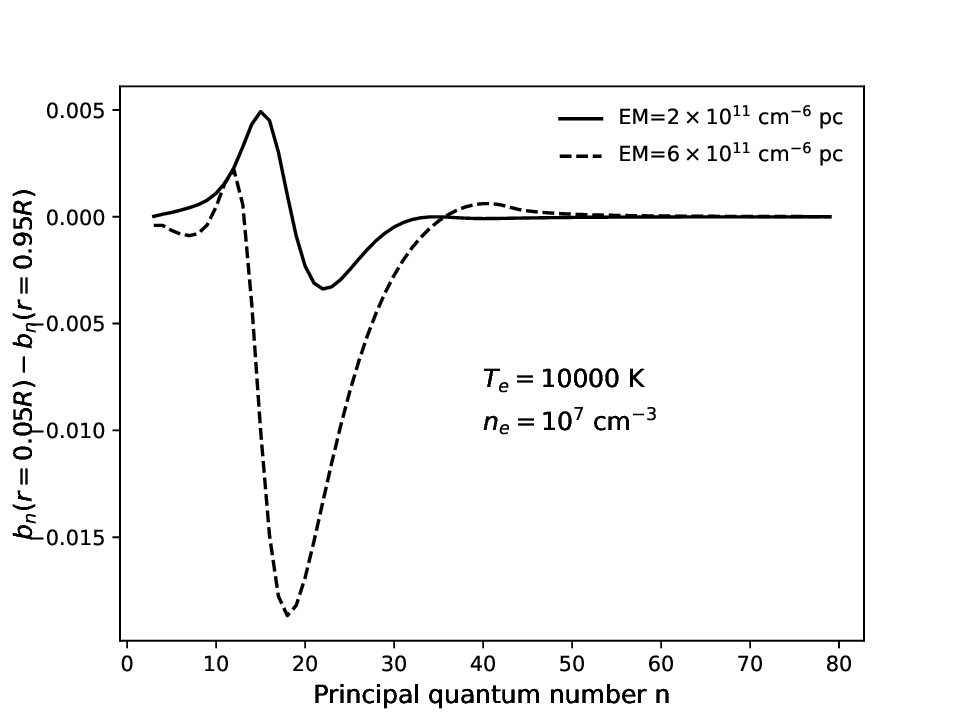}
\caption{The comparison of the calculated departure coefficients $b_n$ with principal quantum number $n$ using the Case B model (blue), the EPA model (red), and the current model with full radiative transfer (black). The dashed and dash-dotted lines respectively indicate the values at the locations 0.05R (center) and 0.95R (boundary) distant from the central massive star in the spherical H II region. $R$ is the radius of a spherical H II region. The electron temperature and density of the H II region are $10000$ K and $10^7$ cm$^{-3}$. The results for $EM=2\times10^{11}$ cm$^{-6}$ pc and $EM=6\times10^{11}$ cm$^{-6}$ pc are shown in the top and middle panels, respectively. The differences between $b_n$ in the center and the boundary of H II regions of the corresponding EMs are plotted in the bottom panel.}\label{fig:departure_EM}
\end{center}
\end{figure}

\begin{figure}
\begin{center}
\includegraphics[scale=0.5]{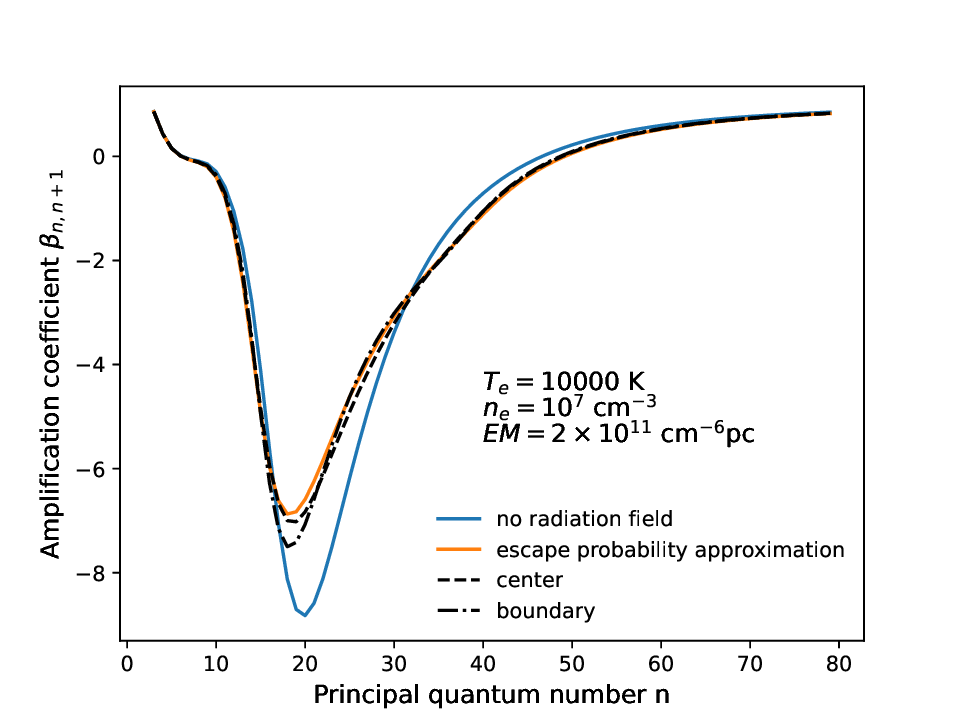}
\includegraphics[scale=0.5]{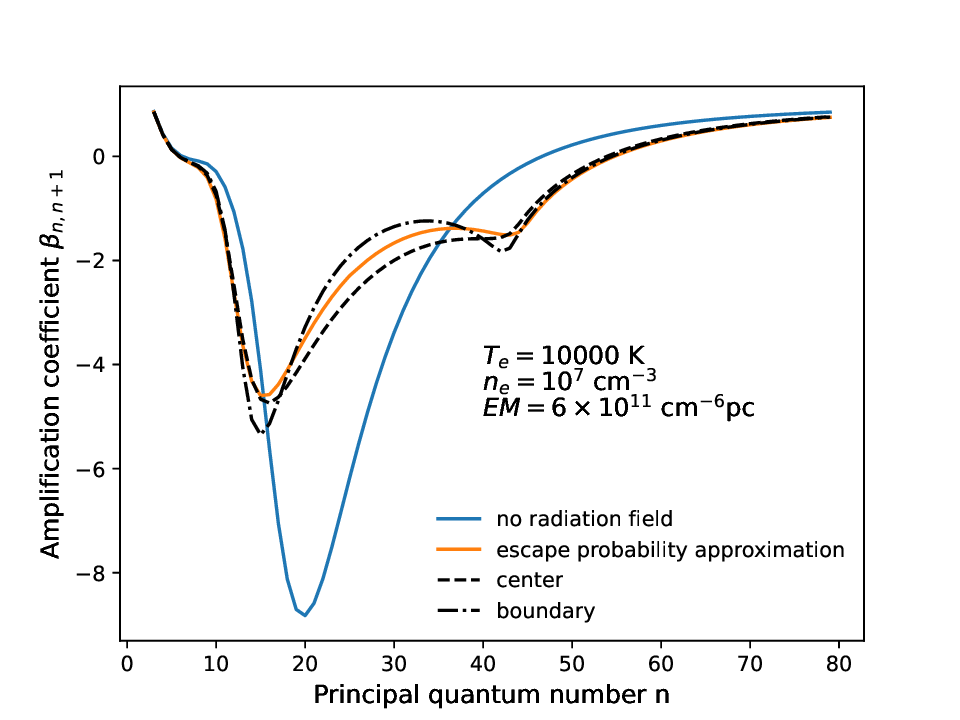}
\includegraphics[scale=0.5]{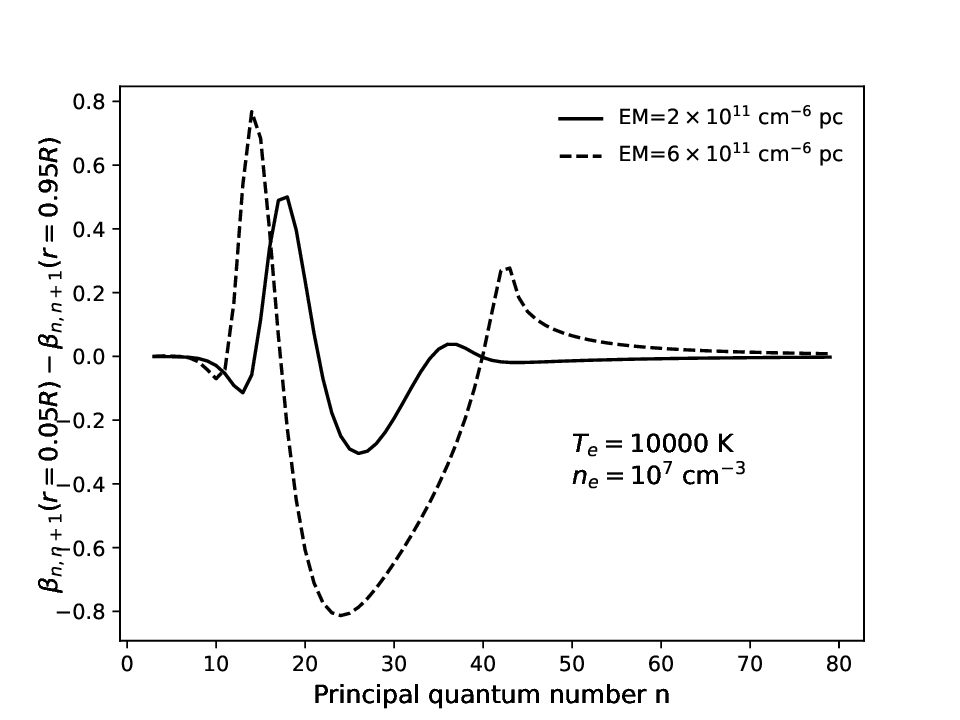}
\caption{The comparison of the calculated amplification coefficients $\beta_{n,n+1}$ using the Case B model, the EPA model, and the current model. The electron temperature and density of the H II region are $10000$ K and $10^7$ cm$^{-3}$. The values of EM are $2\times10^{11}$ cm$^{-6}$ pc and $6\times10^{11}$ cm$^{-6}$ pc for the cases shown in the top and middle panels, respectively. The differences between $\beta_{n,n+1}$ in the center and the boundary of H II regions of the corresponding EMs are plotted in the bottom panel.}\label{fig:beta_EM}
\end{center}
\end{figure}

\begin{figure}
\begin{center}
\includegraphics[scale=0.5]{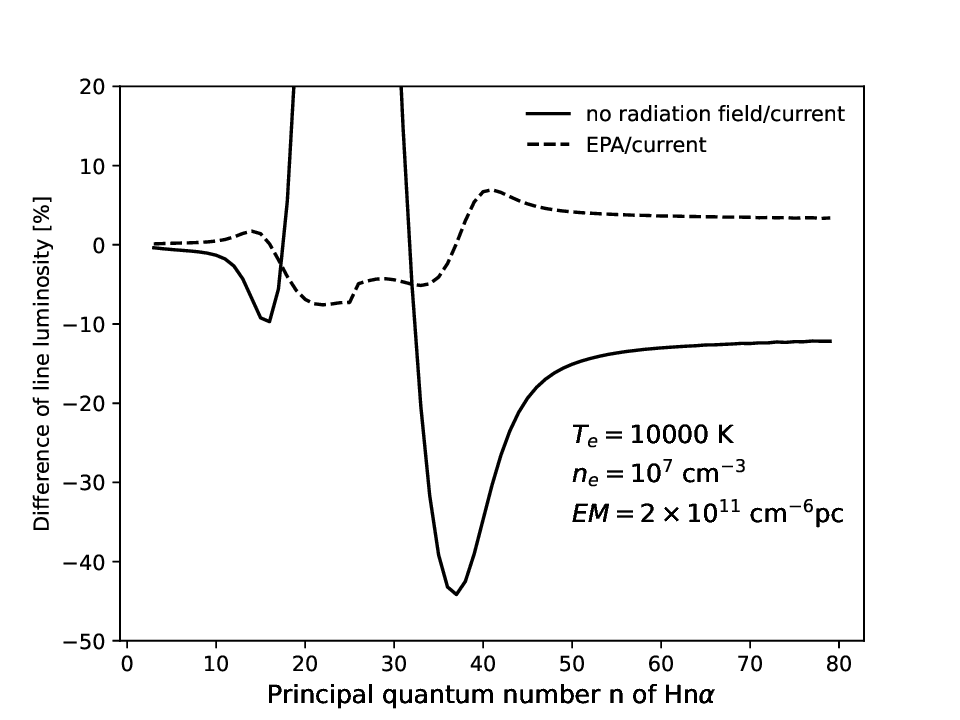}
\includegraphics[scale=0.5]{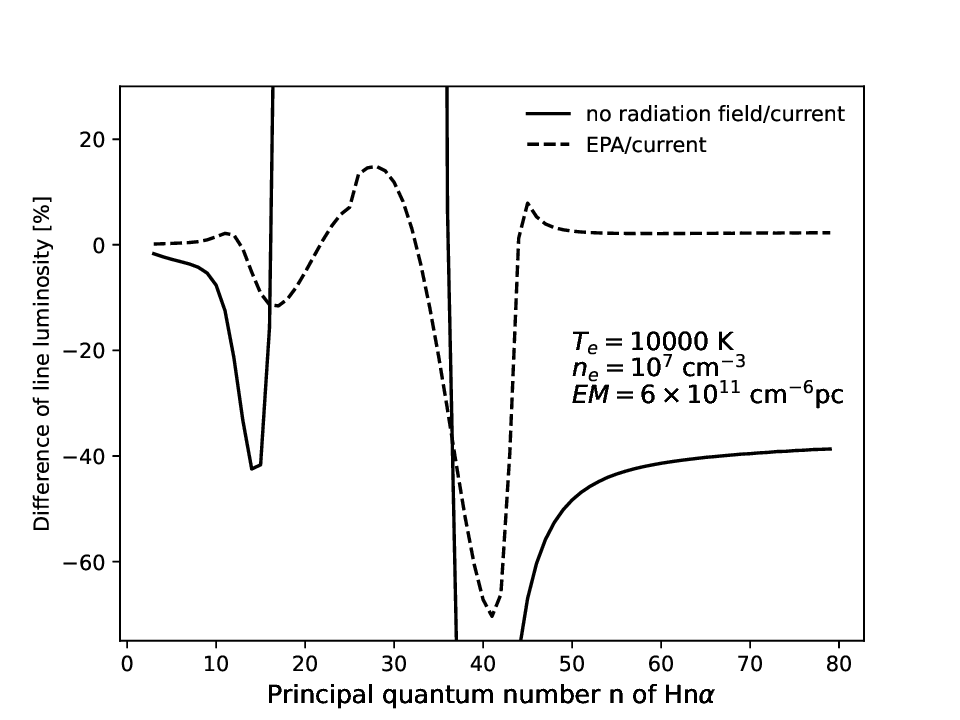}
\caption{The difference between the luminosities of Hn$\alpha$ lines calculated using the Case B model, the EPA model, and the current model. Curves labeled "no radiation field/current" correspond to differential luminosities derived by subtracting our current model values from Case B model results, while those marked "EPA/current" show analogous comparisons between the EPA model and the current model.}\label{fig:flux_EM}
\end{center}
\end{figure}

\begin{figure}
\begin{center}
\includegraphics[scale=0.5]{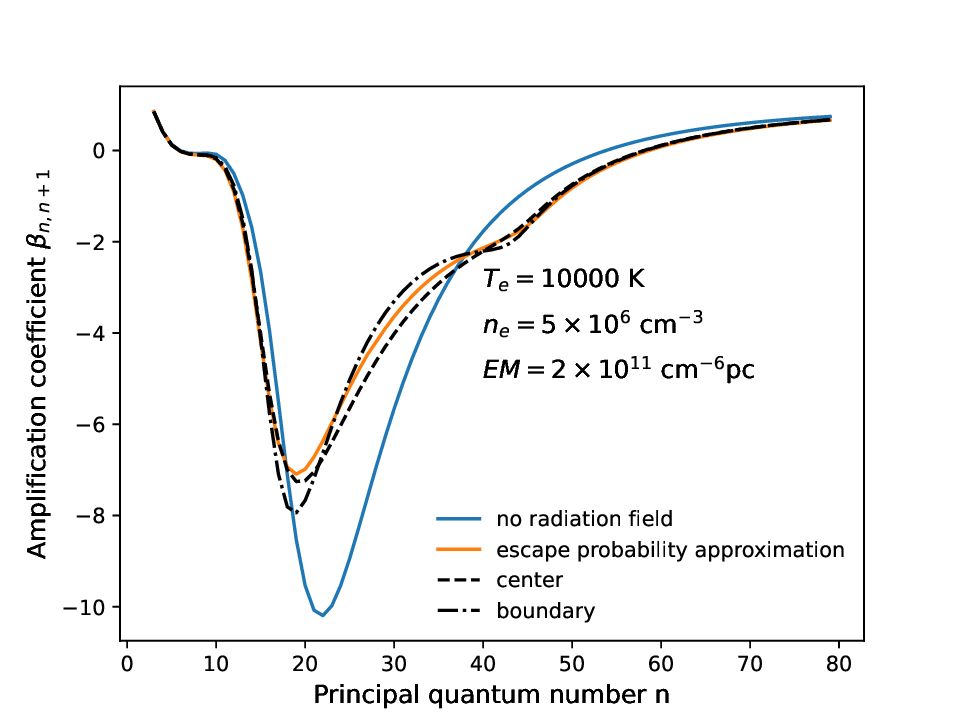}
\includegraphics[scale=0.5]{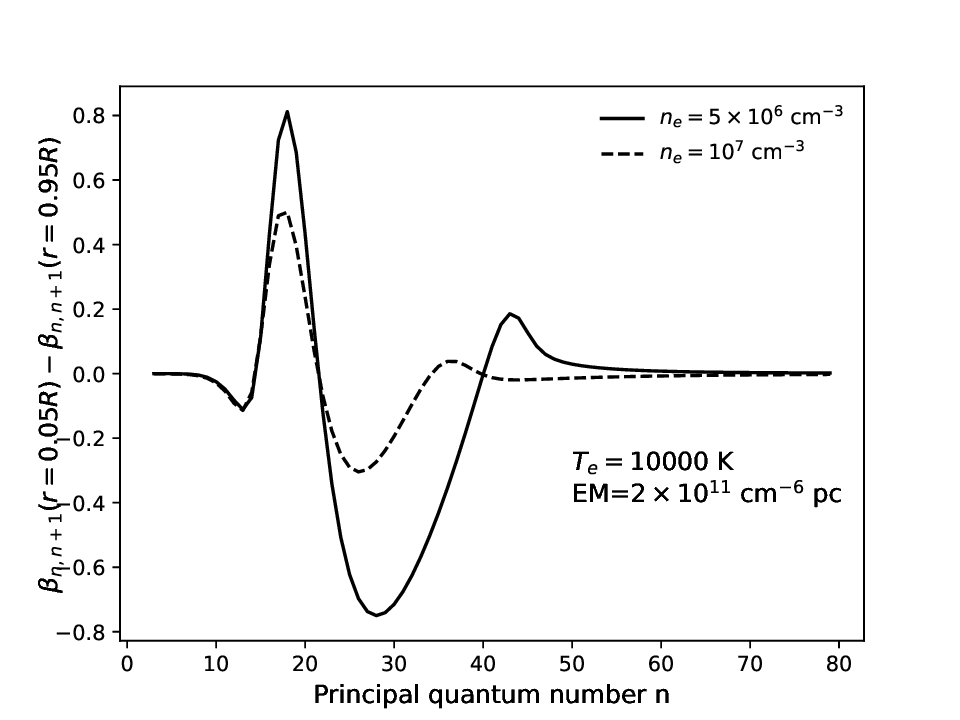}
\caption{The comparison of the calculated amplification coefficients $\beta_{n,n+1}$ using the Case B model, the EPA model, and the current model. The electron temperature and density of the H II region are $10000$ K and $5\times10^6$ cm$^{-3}$. The values of EM are $2\times10^{11}$ cm$^{-6}$ pc. The differences between $\beta_{n,n+1}$ in the center and the boundary of H II regions in this case  are plotted in the bottom panel, in comparison with those in the case of $T_e=10000$ K, $n_e=5\times10^6$ cm$^{-3}$ and EM$=2\times10^{11}$ cm$^{-6}$ pc.}\label{fig:beta_EM_lowdensity}
\end{center}
\end{figure}

\begin{figure}
\begin{center}
\includegraphics[scale=0.5]{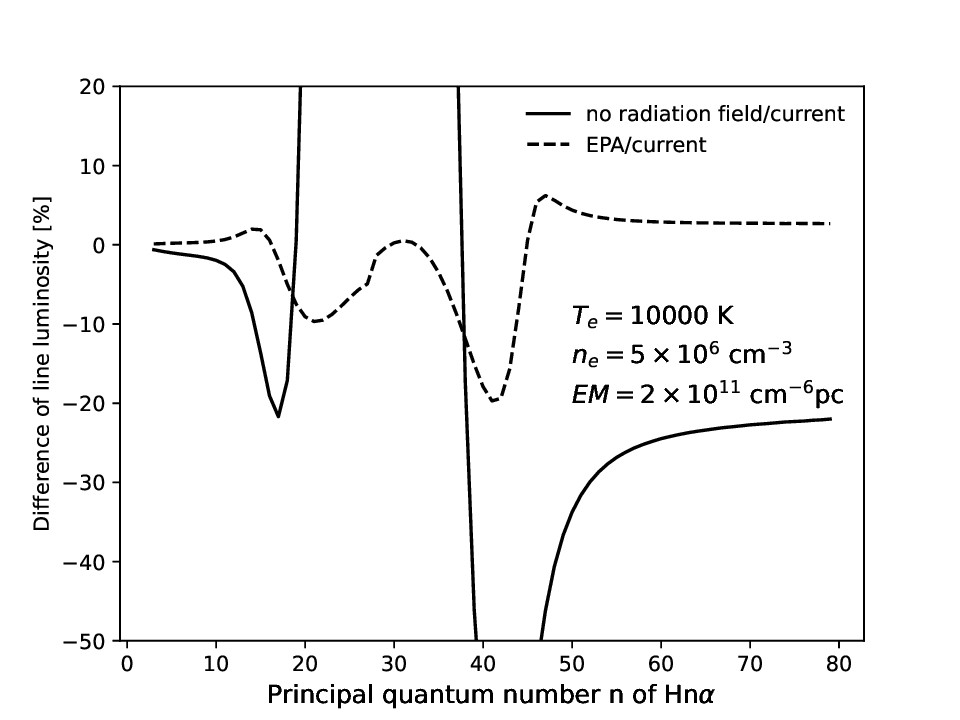}
\caption{The differences between the luminosities of Hn$\alpha$ lines calculated using the Case B model, the EPA model, and the current model. }\label{fig:flux_EM_lowdensity}
\end{center}
\end{figure}

\begin{figure}
\begin{center}
\includegraphics[scale=0.5]{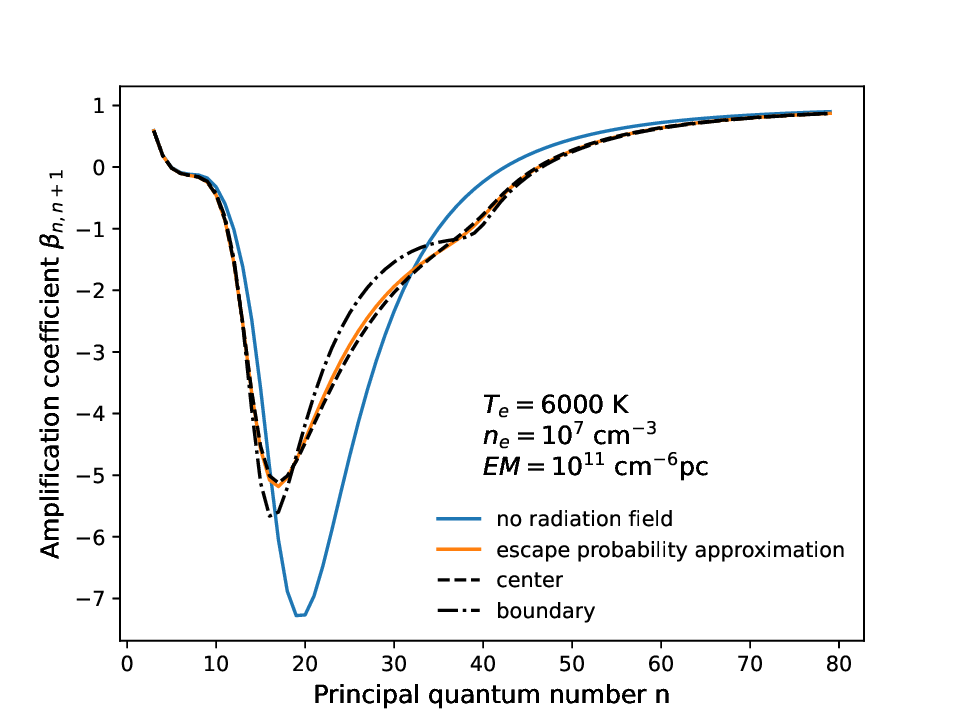}
\includegraphics[scale=0.5]{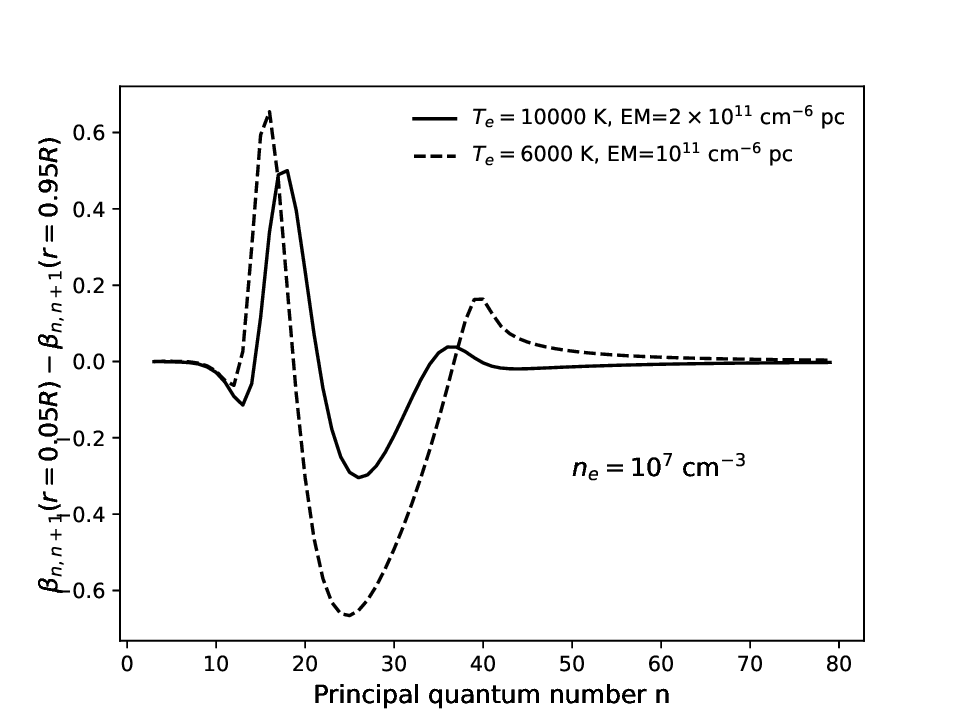}
\includegraphics[scale=0.5]{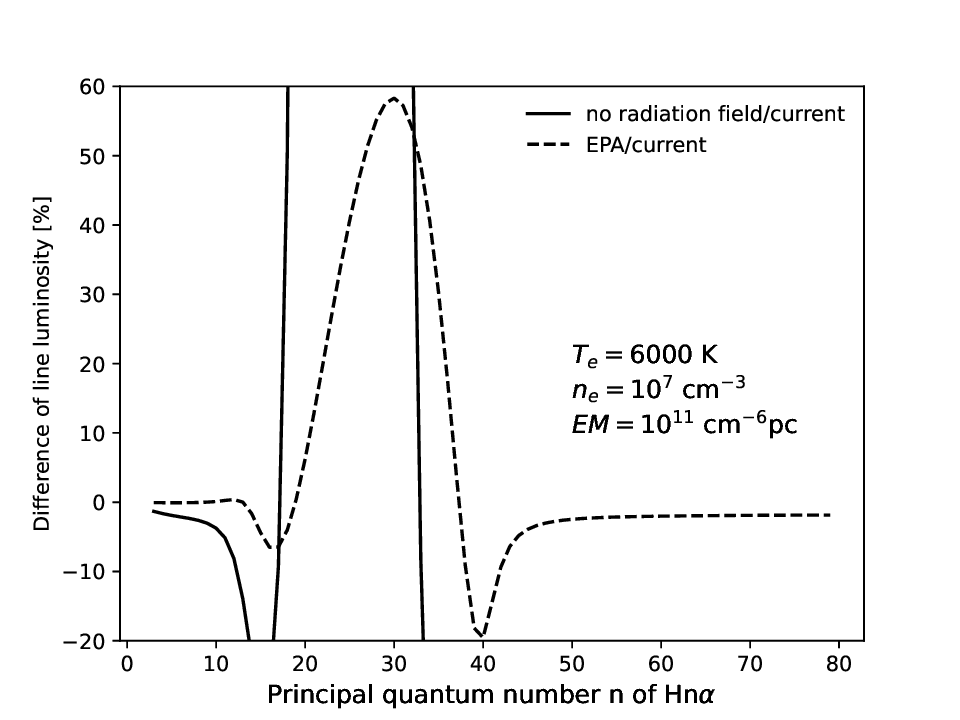}
\caption{The comparison of the calculated amplification coefficients $\beta_{n,n+1}$ using the Case B model, the EPA model, and the current model with full radiative transfer is plotted in the top panel. The electron temperature and density of the H II region are $6000$ K and $10^7$ cm$^{-3}$. The value of EM is $10^{11}$ cm$^{-6}$ pc. The differences between $\beta_{n,n+1}$ in the center and the boundary of H II regions in two cases of identical $n_e$ but different $T_e$ and EM are plotted in the middle panel. In the bottom panel, the differences between the luminosities of Hn$\alpha$ lines calculated using the Case B model, the EPA model and the current model for the case of $T_e=6000$ K, $n_e=10^7$ cm$^{-3}$, and EM$=10^{11}$ cm$^{-6}$ pc are presented.}\label{fig:beta_EM_lowtemperature}
\end{center}
\end{figure}

\begin{figure}
\begin{center}
\includegraphics[scale=0.5]{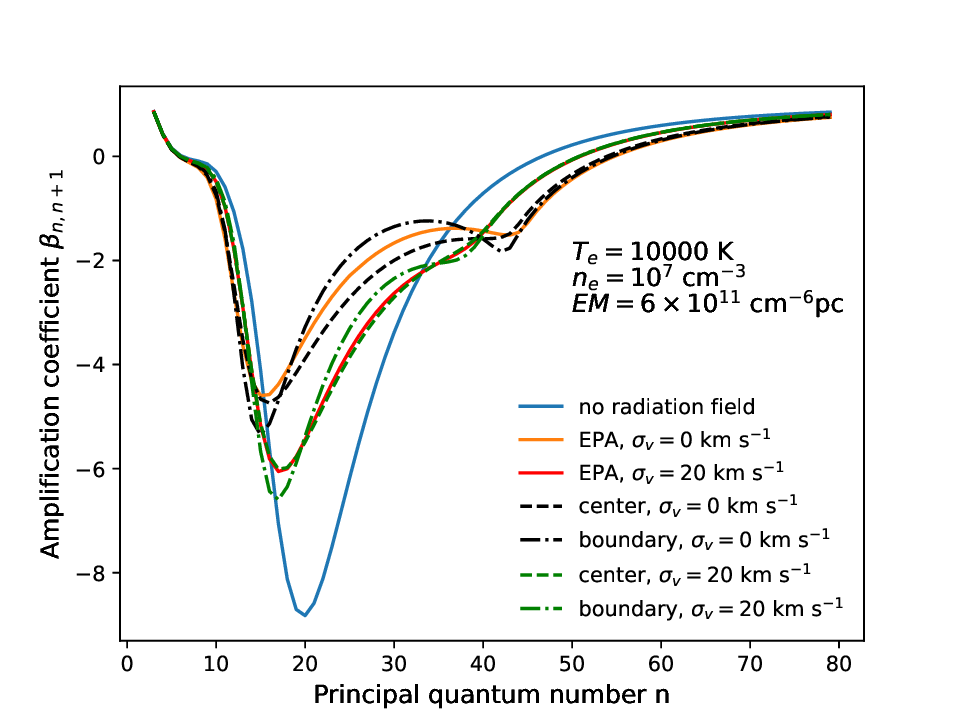}
\includegraphics[scale=0.5]{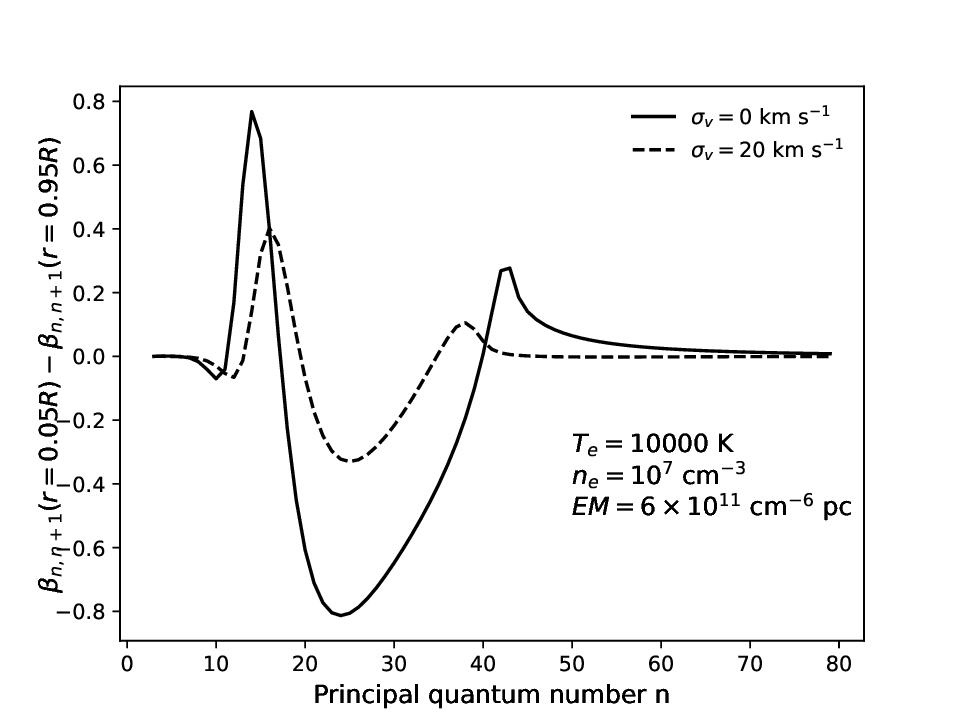}
\includegraphics[scale=0.5]{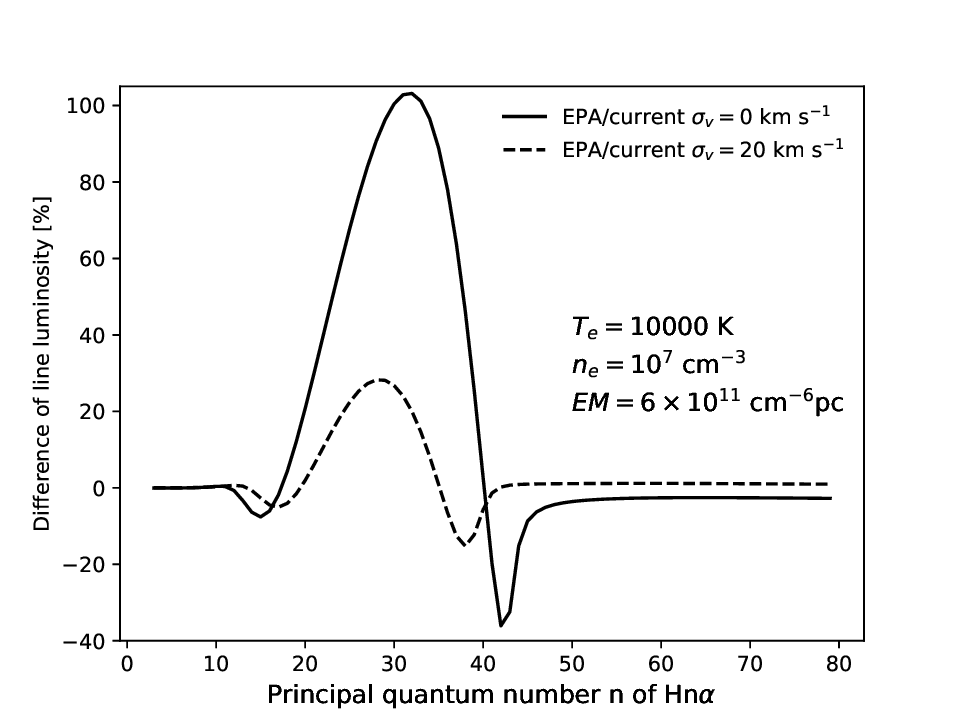}
\caption{The comparison of the calculated amplification coefficients $\beta_{n,n+1}$ using the EPA model and the current model with full radiative transfer is plotted in the top panel. The corresponding comparisons of the $\beta_{n,n+1}$ in the center and boundary of H II regions, and the line luminosities are shown in the middle and bottom panels, respectively. The electron temperature, density and EM are $10000$ K, $10^7$ cm$^{-3}$, and $6\times10^{11}$ cm$^{-6}$ pc, respectively. The $\sigma_v$ is 0 km s$^{-1}$ and 20 km s$^{-1}$ in the two cases.}\label{fig:beta_sigmav}
\end{center}
\end{figure}

\begin{figure*}
\begin{center}
\includegraphics[scale=0.35]{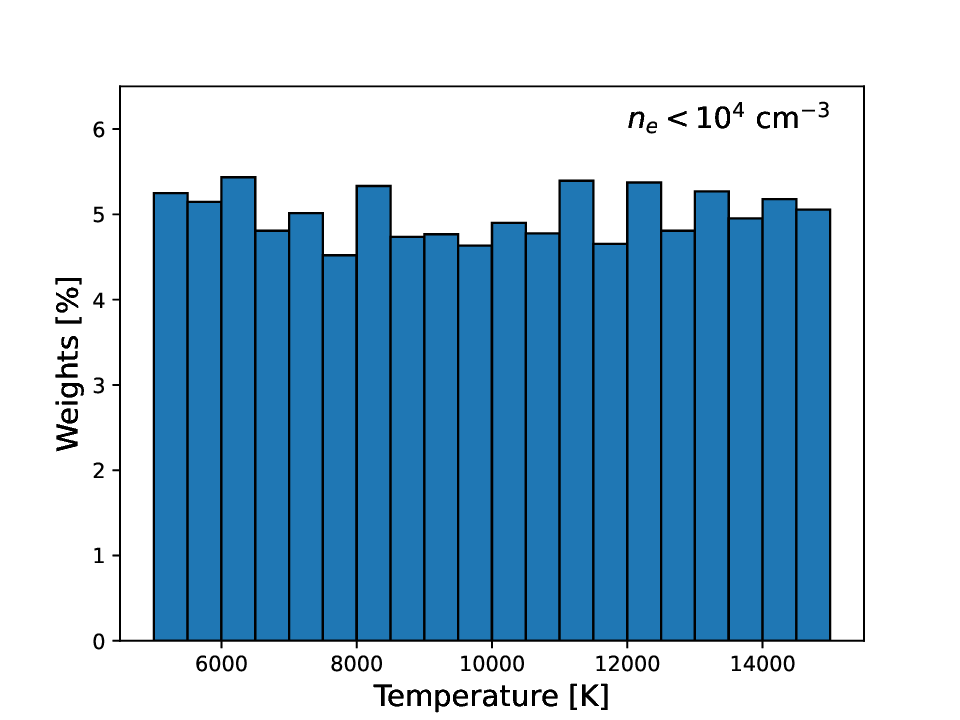}
\includegraphics[scale=0.35]{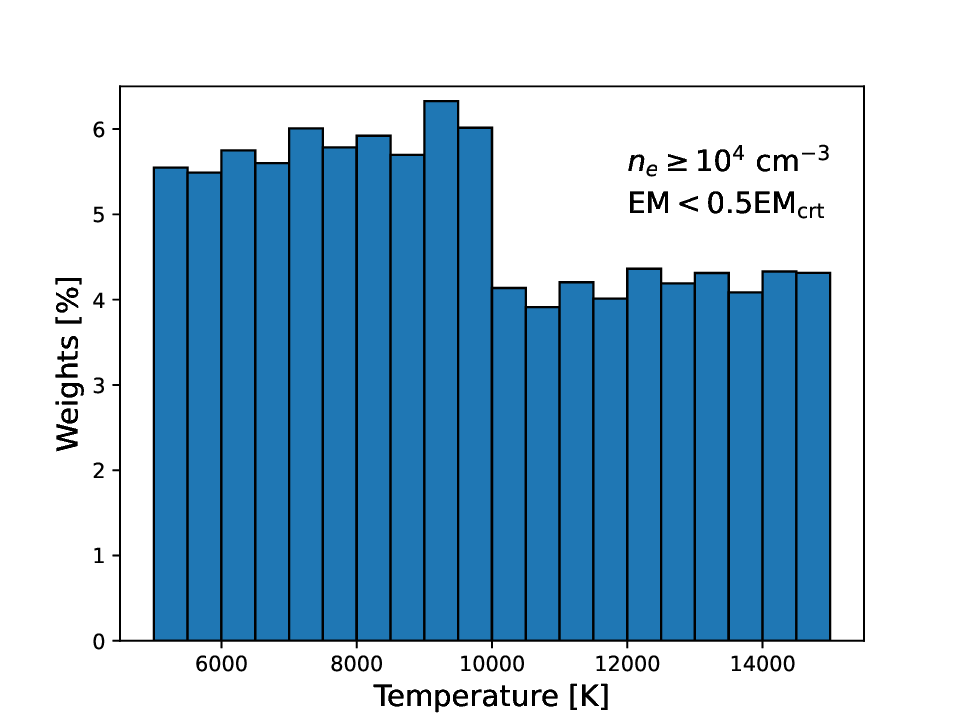}
\includegraphics[scale=0.35]{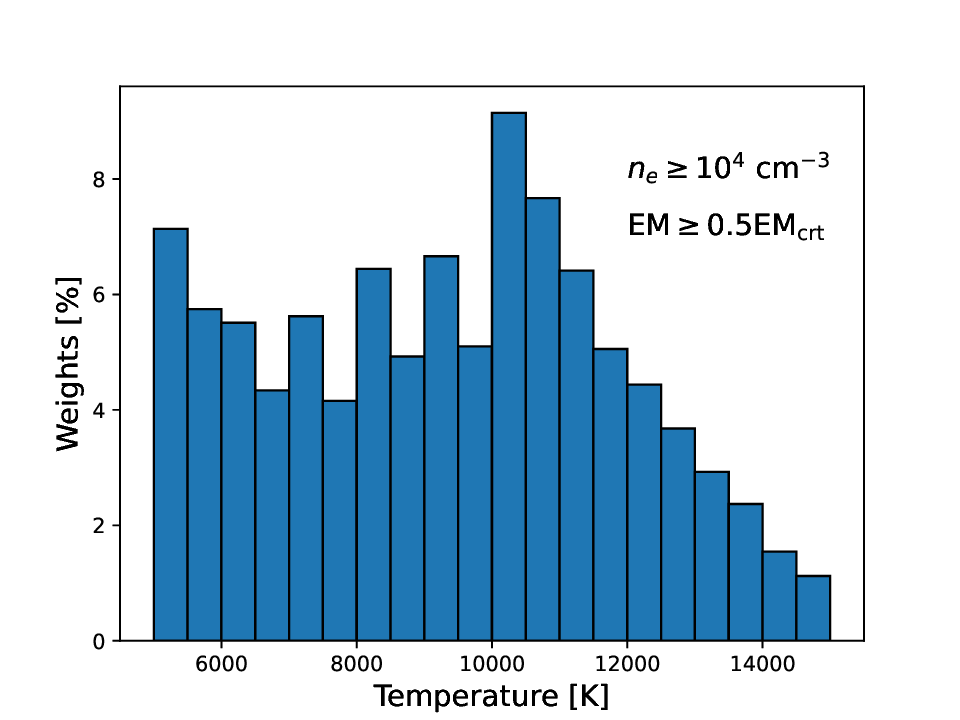}
\includegraphics[scale=0.35]{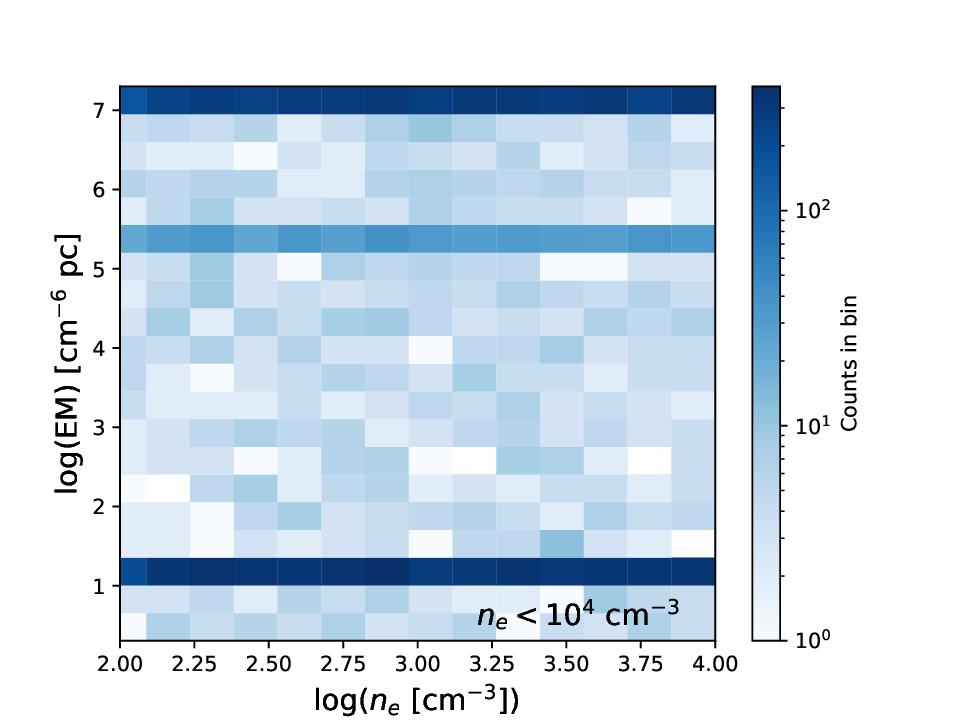}
\includegraphics[scale=0.35]{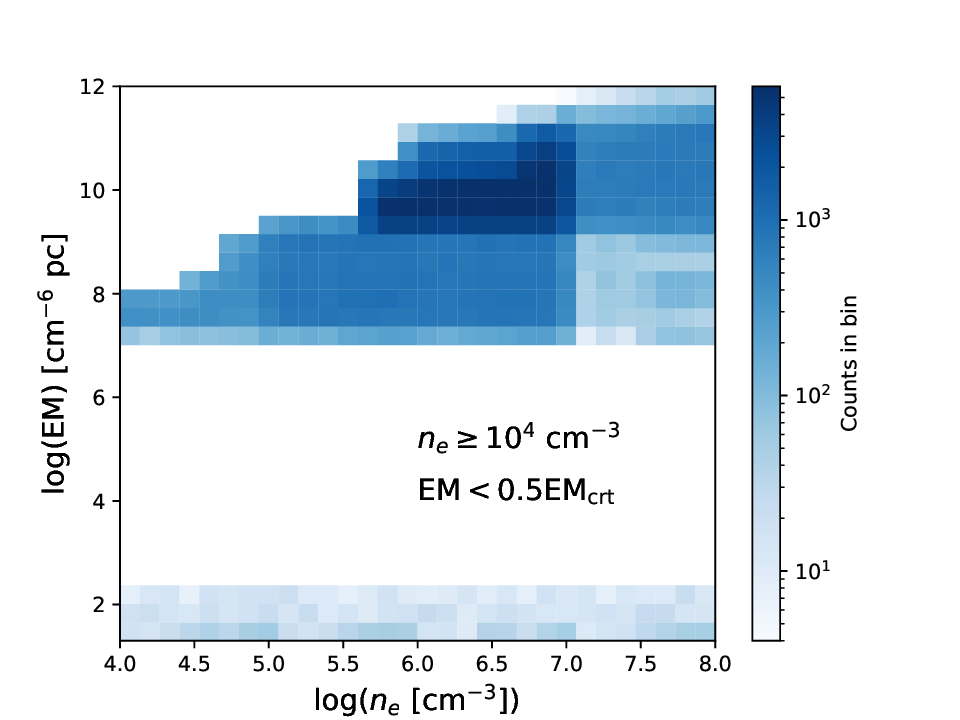}
\includegraphics[scale=0.35]{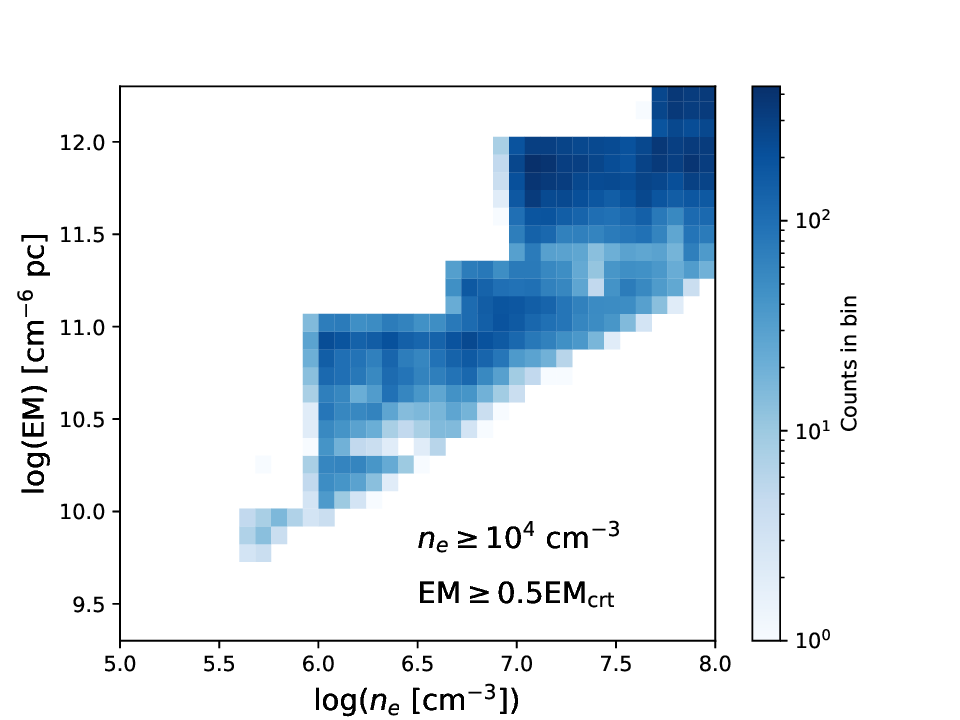}
\includegraphics[scale=0.35]{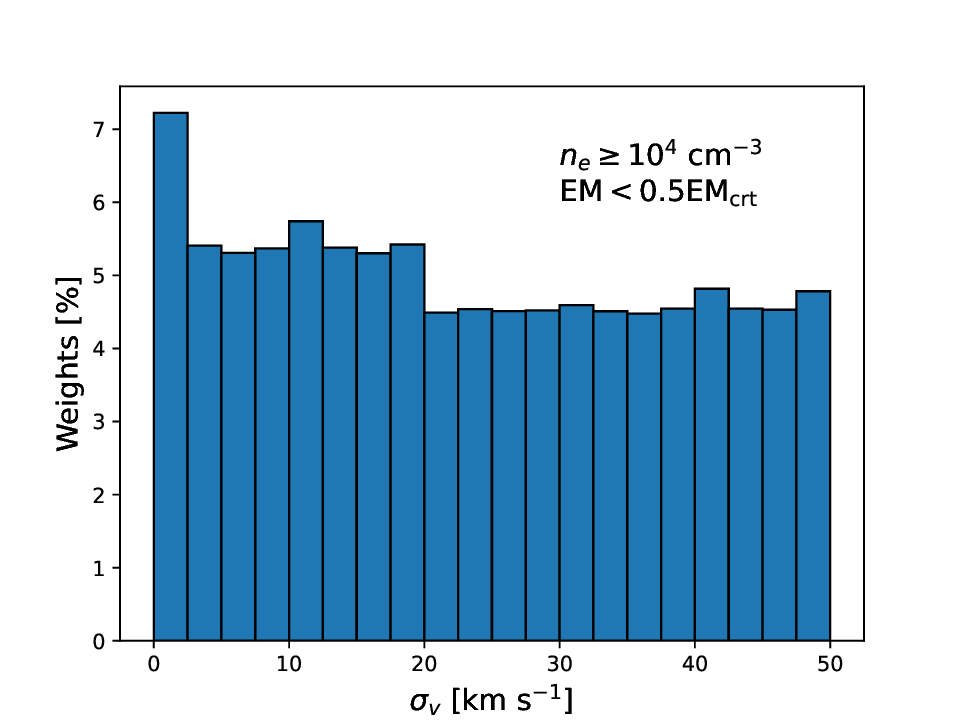}
\includegraphics[scale=0.35]{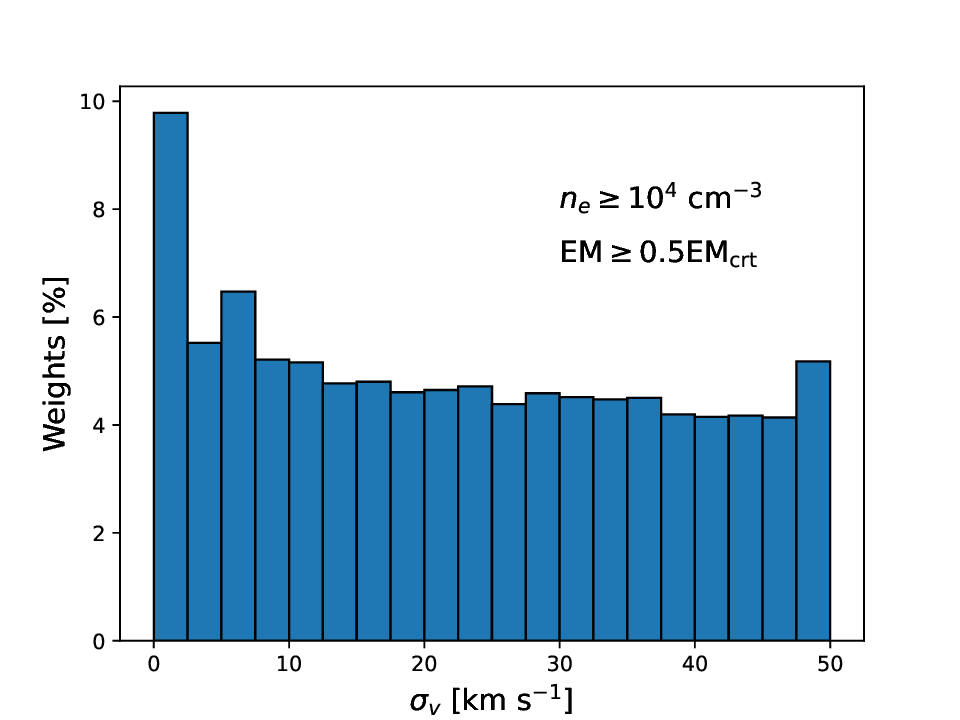}
\caption{The temperature, electron density-EM, and $\sigma_v$ distributions. The temperature distributions in the three subsets of samples are presented in the top panels. The density-EM distributions are shown in the middle panels. The distributions of $\sigma_v$ in the two subsets including samples of $n_e\geq10^4$ cm$^{-3}$ are plotted in the bottom panels. }\label{fig:tem_sample}
\end{center}
\end{figure*}

\end{document}